\documentclass[traditabstract]{aa}  
\usepackage{txfonts}
\usepackage{caption}
\usepackage{epstopdf}
\usepackage{tabu}
\usepackage{array,booktabs}
\usepackage[colorlinks=true,linkcolor=bluea,allcolors=blue]{hyperref}
\usepackage{siunitx}
\usepackage[switch]{lineno}
\usepackage{graphicx,longtable,lscape,natbib,amssymb,amssymb,amsmath}
\usepackage{float}
\bibpunct{(}{)}{;}{a}{}{,} 
\newcommand{\ltsima} {$\; \buildrel < \over \sim \;$}  
\newcommand{\gtsima} {$\; \buildrel > \over \sim \;$}  
\newcommand{\lta} {\lower.5ex\hbox{\ltsima}}  
\newcommand{\gta} {\lower.5ex\hbox{\gtsima}}

\newcommand{\fesc} {f$_{esc}$} 
\newcommand{\WHz}{\>{\rm W}\,{\rm Hz}^{-1}}

\newcommand{\kms}{$\rm{\,km \,s}^{-1}$}
\newcommand{\lya}{Ly$\alpha\,$}

\usepackage{color}
\usepackage{xltabular}
\usepackage{longtable}

\usepackage{lineno}
\makeatletter
\renewcommand*\aa@pageof{, page \thepage{} of \pageref*{LastPage}}
\makeatother

\begin{document}

\title{The MUSE view of Lyman-$\alpha$ nebulae in High redshift Radio Galaxies} 

\author{Miguel Coloma Puga \inst{1,2}, Barbara Balmaverde \inst{2}, Alessandro Capetti\inst{2}, Francesco Massaro\inst{1}, Roberto Gilli\inst{3}}
\institute{
  Dipartimento di Fisica, Universit\`a degli Studi di
  Torino, Via Pietro Giuria 1, 10125 (Torino), Italy
   \and
  INAF - Osservatorio
  Astrofisico di Torino, Via Osservatorio 20, I-10025 Pino Torinese,
  Italy
  \and
  Osservatorio di Astrofisica e Scienza dello Spazio di Bologna
  Via Gobetti 93/3, I-40129 Bologna, Italy
  }

  \abstract{We present the results of VLT/MUSE integral field spectroscopic observations of the Ly$\alpha\,$ emission nebulae associated with 11 high redshift ($z \geq 2.9$) radio galaxies (HzRGs) with DEC $<25^{\circ}$. When considering the other nine sources with available archival MUSE data, these observations increase the coverage to half of the currently known HzRGs. For two sources, we are unable to confirm the original identification, as no Ly$\alpha\,$ emission was detected. We produce narrowband images centered on the Ly$\alpha\,$ line, extract their nuclear spectra, map their ionized gas kinematics, and derive the Ly$\alpha\,$ surface brightness profiles (SBPs). The SBPs are generally well reproduced by an exponential law with a typical scale length of $\sim 20-30$ ckpc. We measure emission line ratios, finding most sources in agreement with an AGN origin for their gas ionization, with a single object hinting at strong star formation. Regarding the connection between the radio and ionized gas emission, we find that while the Ly$\alpha\,$ nebulae are preferentially aligned with the direction of the radio emission, there is no clear correlation in terms of size or gas kinematics and only a weak trend connecting their radio and Ly$\alpha\,$ luminosities. The alignment is most likely the result of anisotropic nuclear emission rather than of a direct impact of the jets into the ionized gas.}

  \titlerunning{\lya\ nebulae in high redshift radio galaxies} \authorrunning{Coloma Puga et al.}
\maketitle
  
  \section{Introduction}
\label{intro}

At high redshift, the emission lines commonly used to trace ionized gas (e.g. [O III]$\lambda\lambda$4959,5007, H$\alpha$) are redshifted into the infrared bands, greatly hindering ground-based observations of distant and faint objects. In return, rest-frame UV lines which are unavailable at low redshift due to atmospheric absorption become visible in optical bands; chief among them is Lyman $\alpha$.

\lya\ radiation in galaxies is created as a byproduct of hydrogen recombination; $\sim$60-70\% of the photoionized hydrogen atoms will yield a \lya\ photon as they cascade down to lower energy levels \citep{Cantalupo_2008}. It is a highly resonant transition, with a large cross-section of interaction between \lya photons and the ever-present HI atoms. As such, the \lya\ emission, even when produced in its majority in a small region within a galaxy (e.g., an AGN), will become imprinted with the kinematic information corresponding to the different phases of intervening HI gas it interacted with as it traveled through the interstellar (ISM) and circumgalactic (CGM) media. This results in complex profiles, with strong absorption troughs and tracing multiple emission components.

The \lya\ line profile tends to be more heavily absorbed on the blue side as compared to the red side. This asymmetry is observed in the redshift measurements performed with both \lya\ and nonresonant ionized lines, as the profile centroids are systematically offset by $\sim$400 \kms\ redward from the systemic velocity \citep{steidel2010}. A great amount of effort has been invested in radiative transfer simulations of \lya, considering a plethora of different configurations for the CGM and ISM geometries in order to explain the observed line profiles (e.g. \citealt{Verhamme_2006}, \citealt{Behrens_2014}, \citealt{Gronke_2015} and \citealt{Gronke_2017}).

Studying objects at high redshift allows us to peek into the cosmic noon (z$\sim$2 - 3), which corresponds to an epoch in the universe's history in which the cosmic averages of star formation rate density, black hole accretion and AGN activity were at or very close to their maxima (\citealt{Madau2014,D_Silva_2023,wang2024cosmic}). As such, the effects of AGN in their host galaxies and the large-scale environments surrounding them are expected to be particularly pronounced.

\lya\ has been routinely utilized to trace nebular emission extending up to 100s of kpc from the nucleus of high redshift AGN (see \citealt{Cantalupo_2014,borisova16,Arrigoni_Battaia_2018,Cai_2019}), utilizing the new generation of integral field units (IFU), namely MUSE at the Very Large Telescope and KCWI at the W. M. Keck Observatory. The combination of fluorescent emission due to resonant scattering and the intrinsic brightness of the line \citep{Hayes_2015} enables the observations of these gargantuan gas structures, whereas other ionized tracer cannot be used to probe at such depths.

In particular, high redshift radio galaxies (HzRGs) represent an ideal laboratory in which one can study the interplay between nuclear activity and AGN host galaxies, as the selective obscuration caused by the circumnuclear medium entirely eliminates the need to remove the quasar light, which is essential in type 1 AGN if one wishes to analyze the galactic component. This is a process that, specially at high redshift due to objects appearing increasingly fainter, can generate large uncertainties and loss of valuable information. 

HzRGs are a well studied (\citealt{Villar_Mart_n_2007_a,Villar_Martin_2007_b,Miley_2008,Matsuoka_2009,Nesvadba_2017,Silva_2017,Wang_2023}) but rare class of object, with about $\sim$50 known HzRGs above z=3 \citep{Miley_2008}. They are very massive (\citealt{Rocca_Volmerange_2004,Seymour_2007,De_Breuck_2010}), vigorously star-forming \citep{Seymour_2008,Drouart_2014} and reside in overdense environments \citep{Venemans_2006,Hatch_2010,Wylezalek_2013,Uchiyama_2022}.

\citet{Wang_2023} presented the results of the MUSE observations of eight HzRGs, selected from the list compiled by \citet{Miley_2008} and obtained as part of several observing programs. These sources were selected because in previous observations they had been found to be associated with large \lya\ nebulae, all extended over more than 10\arcsec. In this paper we present the results of MUSE observations of an additional 11 HzRGs accessible for observations at the VLT ($\delta < 25^\circ$), and \lya\ visible in their MUSE spectra (z$\geq$ 2.9) so as to gain a more general view of the properties of \lya\ nebulae around these sources.

\begin{table*}
\caption{Main properties of the HzRGs sample and observations log.}
    \label{tab1}
    \centering
    \begin{tabular}{l | l l l l r l l}
Name & RA & Dec. & z & Program & Exp. time (s) & Reference & P$_{500}$.\\
\hline
NVSS~J213238-335318 & 21:32:39.0 & $-$33:53:19 & 2.900 & 60.A-9334(A) & 6510 &           & 28.60\\
MRC~0943-242        & 09:45:32.8 & $-$24:28:50 & 2.922 & 096.B-0752(A) & 15146 & Wang23  & 28.62\\
NVSS~J151020-352803 & 15:10:20.8 & $-$35:28:03 & 2.937 & 108.22FU.001 & 5226 & This work & 28.35\\
TXS~0952-217        & 09:54:29.5 & $-$21:56:53 & 2.950 & 108.22FU.001 & 2607 & This work & 28.08\\
TN~J1112-2948       & 11:12:23.9 & $-$29:48:06 & 3.090 & 108.22FU.001 & 2595 & This work & 28.76\\
NVSS~J095751-213321 & 09:57:51.3 & $-$21:33:21 & 3.126 & 108.22FU.001 & 2605 & This work & 28.11\\
MRC~0316-257        & 03:18:12.1 & $-$25:35:10 & 3.142 & 094.B-0699(A) & 15264 & Wang23  & 28.95\\
MRC~0251-273        & 02:53:16.7 & $-$27:09:10 & 3.160 & 108.22FU.001 & 2589 & This work & 28.54\\
NVSS~J230123-364656 & 23:01:23.5 & $-$36:46:56 & 3.220 & not observed & & ---            & 28.34\\
NVSS~J232100-360223 & 23:21:00.6 & $-$36:02:25 & 3.320 & not observed  & & ---           & 28.22\\
NVSS~J094724-210505 & 09:47:24.5 & $-$21:05:06 & 3.377 & 108.22FU.001 & 2583  & This work& 27.43\\ 
NVSS~J095438-210425 & 09:54:38.4 & $-$21:04:25 & 3.431 & 108.22FU.001 & 2607 & This work & 28.13\\ 
NVSS~J231402-372925 & 23:14:02.4 & $-$37:29:27 & 3.450 &not observed &             & --- & 28.83\\
TN~J0205+2242       & 02:05:10.7 & $+$22:42:50 & 3.506 & 096.B-0752(B) & 15264 & Wang23  & 28.46\\
TN~J0121+1320       & 01:21:42.7 & $+$13:20:58 & 3.517 & 097.B-0323(C) & 19088 & Wang23  & 28.49\\
4C+03.24 (4C~1243+036)  & 12:45:38.4 & $+$03:23:20 & 3.560 &60.A-9100(G) & 4500 & Wang23 & 29.23\\
MG~2141+192 (4C +19.71) & 21:44:07.5 & $+$19:29:15 & 3.592 &097.B-0323(B)& 20984& Wang23 & 29.08\\
TN~J1049-1258       & 10:49:06.2 & $-$12:58:19 & 3.697 & 108.22FU.001 & 2601 & This work & 28.94\\
TN~J2007-1316       & 20:07:53.2 & $-$13:16:44 & 3.837 &not observed & & ---             & 29.13\\
NVSS~J231727-352606 & 23:17:27.4 & $-$35:26:07 & 3.874 &not observed & & ---             & 28.71\\
NVSS~J021308-322338 & 02:13:08.0 & $-$32:23:40 & 3.976 & 108.22FU.001 & 2596 & This work & 28.40\\
TN~J1338-1942       & 13:38:26.1 & $-$19:42:30 & 4.105 &60.A-9100(B) & 24840 & Wang23    & 28.70\\
TN~J1123-2154       & 11:23:10.2 & $-$21:54:05 & 4.109 & 108.22FU.001 & 3929 & This work & 28.45\\
RC~J0311+0507 4C 04.11 & 03:11:48.0 & $+$05:08:03 & 4.514 &096.B-0752(F) &15264 & Wang23 & 29.49\\
TN~J0924-2201       & 09:24:19.9 & $-$22:01:42 & 5.197 & 108.22FU.001& 2610 & This work  & 28.94\\ 
\hline
\end{tabular}
\tablefoot{List of the 25 HzRGs in \citet{Miley_2008} with $z>2.9$ and $\delta < 25^\circ$ in order of increasing redshift. We report their coordinates and redshifts, the program of MUSE observations, the exposure time, and the reference where the data are published. The last column includes the logarithm of the radio luminosity at the rest frame frequency of 500 MHz, also taken from \citet{Miley_2008}.}
\end{table*}

The paper is organized as follows: in Sect. \ref{observations} we present the sample and describe the observations and the data reduction. In Sect. \ref{results} we present our results divided in the following subsections: (\ref{SBP}) presents the radial distribution of the \lya\ emission, (\ref{vmaps}) the gas kinematics and (\ref{lineratios}) the emission line ratios. The connection between the radio and \lya\ emission is addressed in Sect. \ref{radiovslya}. In Sect. \ref{summary} we summarize the main findings and draw our conclusions.  

We adopt the following set of cosmological parameters: H$_{0}=69.7$ km s$^{-1}$ Mpc$^{-1}$ and $\Omega_{m}=0.286$ (\citealt{bennett14}).

\section{The sample and the MUSE observations}
\label{observations}

The sample selection was inherently limited due to the relative scarcity of known high redshift radio galaxies. From the aforementioned list of HzRGs compiled by \citet{Miley_2008}, we first limit our selection to the 47 sources with z$\geq$ 2.9, which corresponds to the \lya\ emission line cutoff in the MUSE spectral coverage. Additionally, these sources are further restricted to those which are easily observable from the VLT, i.e., with $\delta < 25^\circ$. This results in 25 sources, which represents about half of all known HzRGs at z$\geq$ 2.9. From this list we drop the nine sources for which MUSE data were already available; the analysis of the MUSE observations of said sources was presented for all but one object by \citet{Wang_2023}. Five sources from the original list were not observed due to scheduling constraints. The remaining 11 sources were observed as part of ESO program 108.22FU.001(PI B. Balmaverde). Four separate observations were performed, between which the telescope was rotated 90$^\circ$ to reject cosmic rays, for a total of $\sim$2600s of exposure time at minimum. We used the European Southern Observatory (ESO) MUSE pipeline (version 2.8.7) to obtain a fully reduced and calibrated data cube \citep{Weilbacher20}. The seeing was in the range 0\farcs5 - 1\farcs2, with a median value of 0\farcs71. 

In Table \ref{tab1} we list all 25 sources with $z \geq$ 2.9, in order of increasing redshift, and provide the record of their observations. Overall, the archival MUSE data have significantly longer exposure times (with an average of 15200 s) than those obtained with our program (with an average of 2960 s), a factor of $\sim$5.

We corrected the astrometry of the MUSE data using the position of several sources visible in the PanSTARRS images \citep{Chambers_2016} (or, when these are not available, from the Digitized Sky Survey\footnote{Available at http://archive.eso.org/dss/dss.}) and present in the MUSE field of view. This procedure enables us to align the MUSE data with the radio data derived from the images of the Karl G. Jansky Very Large Array Sky Survey (VLASS \citealt{lacy20}) obtained at a frequency of 3 GHz with a spatial resolution of 2\farcs5, with an accuracy of $\lesssim 0\farcs4$.

\section{Results}
\label{results}

In two of the 11 sources observed, NVSS J021308-322338 and TN J1123-2154, \lya\ emission was not detected in the MUSE data. The 3$\sigma$ upper bound of their \lya\ flux is well below the values reported in literature measurements. Therefore, we cannot confirm the identification of these two sources as HzRGs. They are consequently dropped from the sample and excluded from the remaining analysis in this study. More details are given in \hyperref[appendixC]{Appendix C}. 

In Fig. \ref{fig:j151nuc2} we present as an example the main results obtained from the analysis of the MUSE data for NVSS J151020-352803.
The remaining sources are shown in \hyperref[appendixA]{Appendix A}, Fig. \ref{fig:j151nuc} through \ref{fig:j092nuc}. For each source we present the nuclear spectrum, extracted using an 1\farcs2 radius aperture ($\sim$ 8-10 kpc) centered on the pixel with the brightest \lya\ emission. Whenever additional lines (namely  NV$\lambda\lambda$1238,1242, CIV$\lambda\lambda$1548,1550, HeII$\lambda$1640, and CIII]$\lambda$1908) are observed, they are also presented along \lya. For the objects in which HeII is detected, its centroid is used as a velocity reference, as it is a more robust measurement of systemic redshift than \lya. We also present 20\arcsec$\times$20\arcsec\ images for each source, created using a wavelength range spanning the FWHM of the total line profile and centered on the peak of the \lya\ emission. For the sources in which the radio emission is extended, we over-plot the radio contours onto the \lya\ image. The \lya\ brightness profile extracted in circular annuli of increasing radius (see Sect. \ref{SBP}) are also shown. Finally, we present spatially resolved maps of velocity and line width (see Sect. \ref{vmaps}). In \hyperref[appendixB]{Appendix B} we provide notes on the individual sources.

\begin{figure*}
    \includegraphics[width=\textwidth]{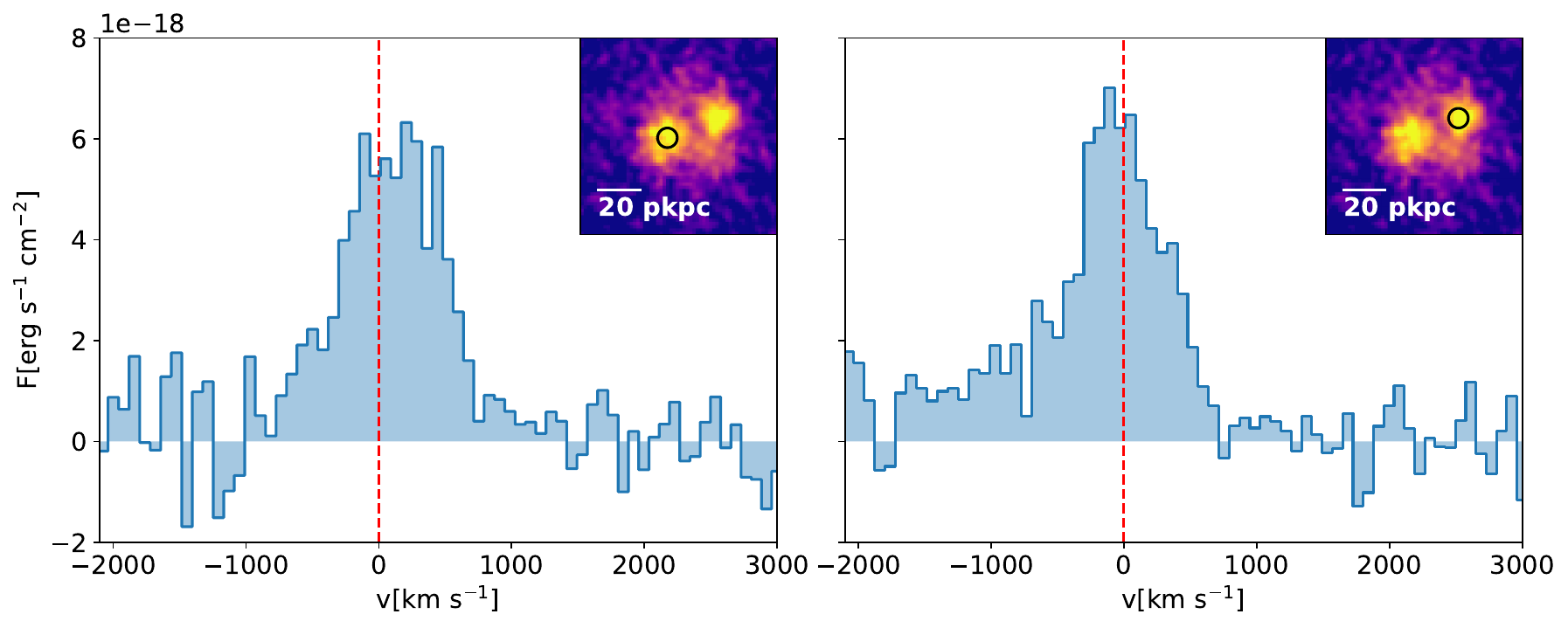} \\
    \includegraphics[width=.4\textwidth]{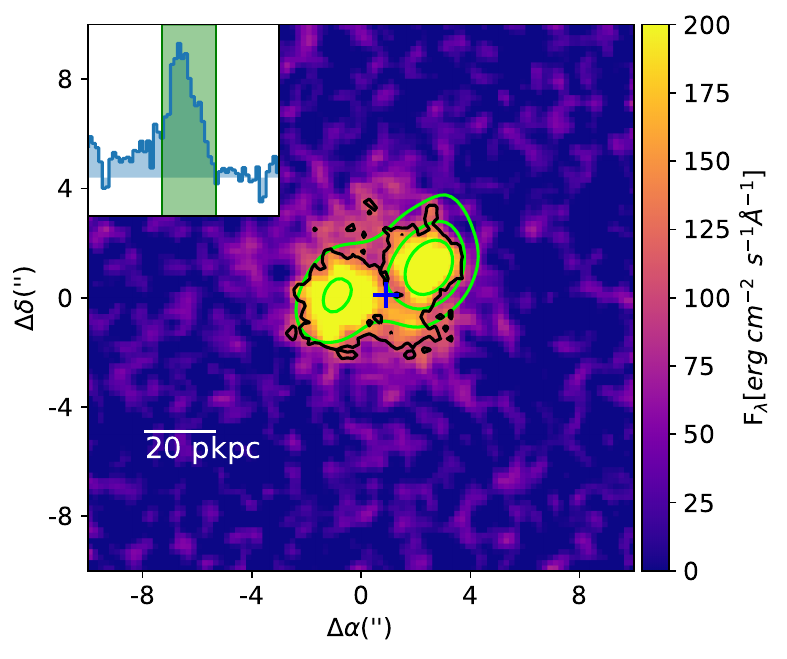}
    \hspace{1cm}
    \includegraphics[width=.5\textwidth]{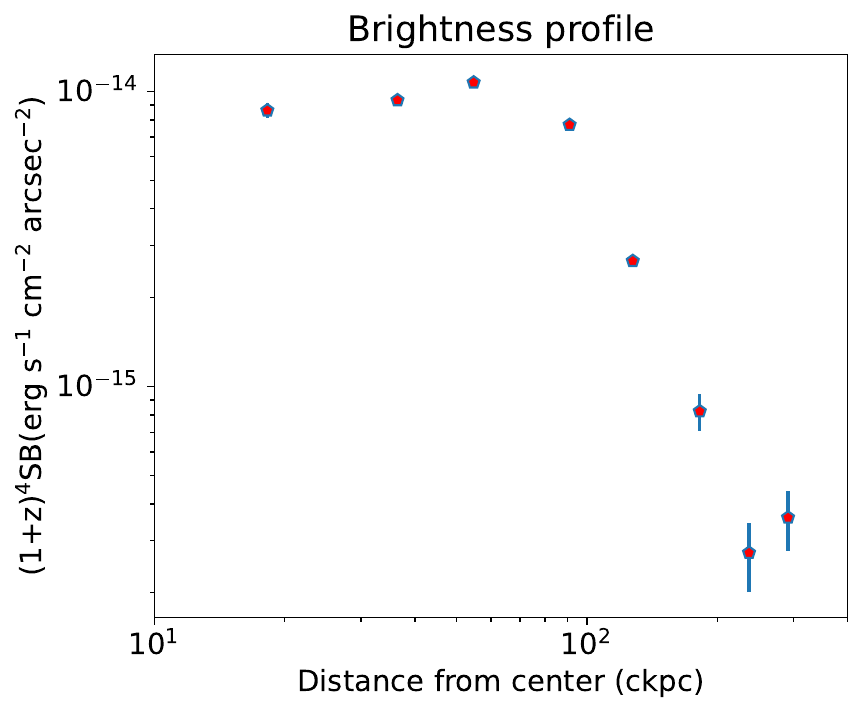}\\
    \includegraphics[width=\textwidth]{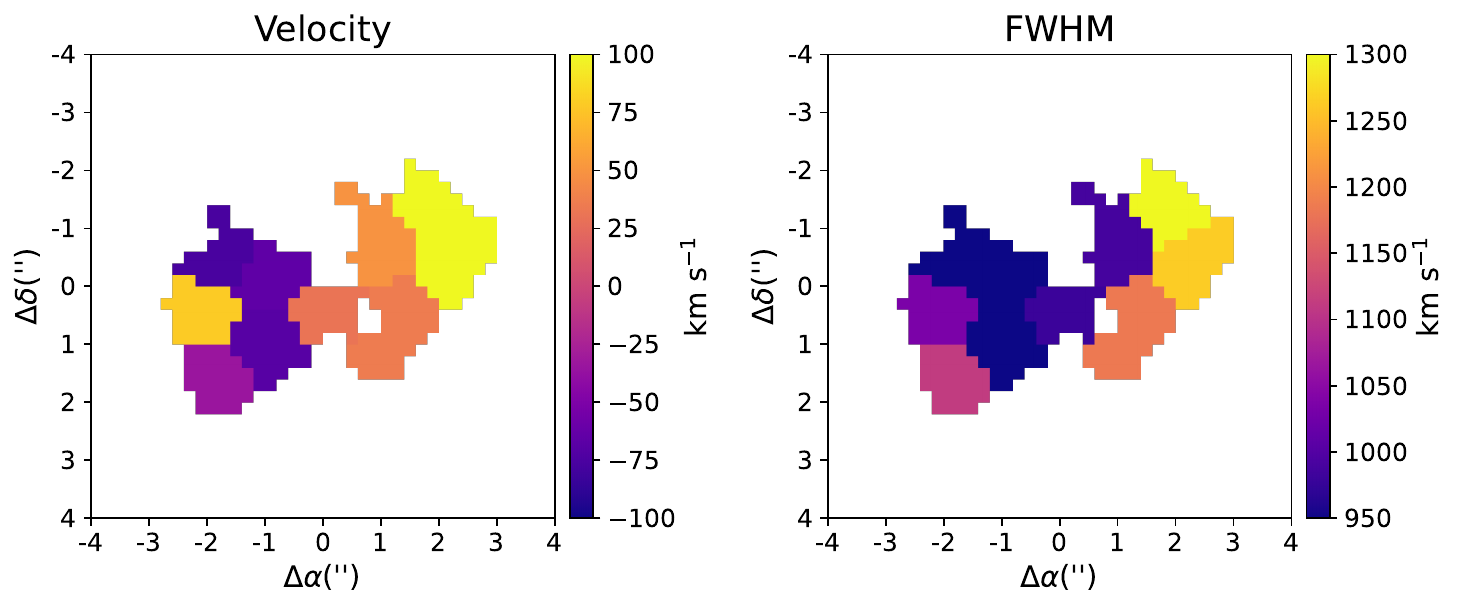}
    \caption{NVSS J151020-352803. Top: spectra extracted from the two \lya\ regions (shown as the black circles in the narrowband images in the insets). The dashed red line represents the systemic redshift, in this case, taken from the \lya\ centroid. Middle: on the left is shown the \lya\ image extracted from the spectral region shown in the inset with superposed radio contours obtained from the VLASS image. The black contours mark the 2$\sigma$ flux limits, and the blue cross is the point from which the brightness profile is extracted. On the right is shown the surface brightness profile. Bottom: Voronoi-tessellated maps of velocity and FWHM of the \lya\ line. }
    \label{fig:j151nuc2}
\end{figure*}

\subsection{Radial distribution of the \texorpdfstring{\lya}{} emission}
\label{SBP}

We produced \lya\ surface brightness profiles (SBPs) extracted in circular annuli of increasing radius. The extraction regions start as a 1-pixel radius circular aperture, and the interval between the radii used to create each annulus increases as they draw away from the center of emission, to increase the total collecting area and decrease noise in zones of fainter nebular emission. The manner in which these intervals are defined ensures that the inner regions, the extended emission and the sky background are all well sampled in a logarithmic scale, as they will be fit with either an exponential or a power law. The sizes of the annuli are maintained across all sources instead of modifying them case by case; this is done in order to have a similar spatial sampling for all of our sources, given that the angular scale changes little within the redshift interval of interest. The procedure is as follows: within every annulus, we sum the relevant spectral pixels from each spaxel to generate a representative spectrum. The continuum is modeled with a single-degree polynomial and removed;  the \lya\ emission is then measured in each annulus by direct integration of the line profile.

We did not attempt to account for the effects of absorption in the \lya\ profile due to the inherent limitations in our sample given the relatively low surface brightnesses and S/N of the sources. Performing the absorption correction of the spectrum in each annulus is unfeasible considering that 1) the outermost annuli would be entirely impossible to correct and we would have to limit ourselves to the highest surface brightness regions 2) even in the central brighter regions, each annulus would contain emission from antipodal regions. This is because both the number of absorbers and the depth of the troughs they produce have been observed to change dramatically in kpc scales (see, e.g., \citealt{puga202420}), making this approach unfeasible. 

\begin{figure}
    \centering
    \includegraphics[width=0.45\textwidth]{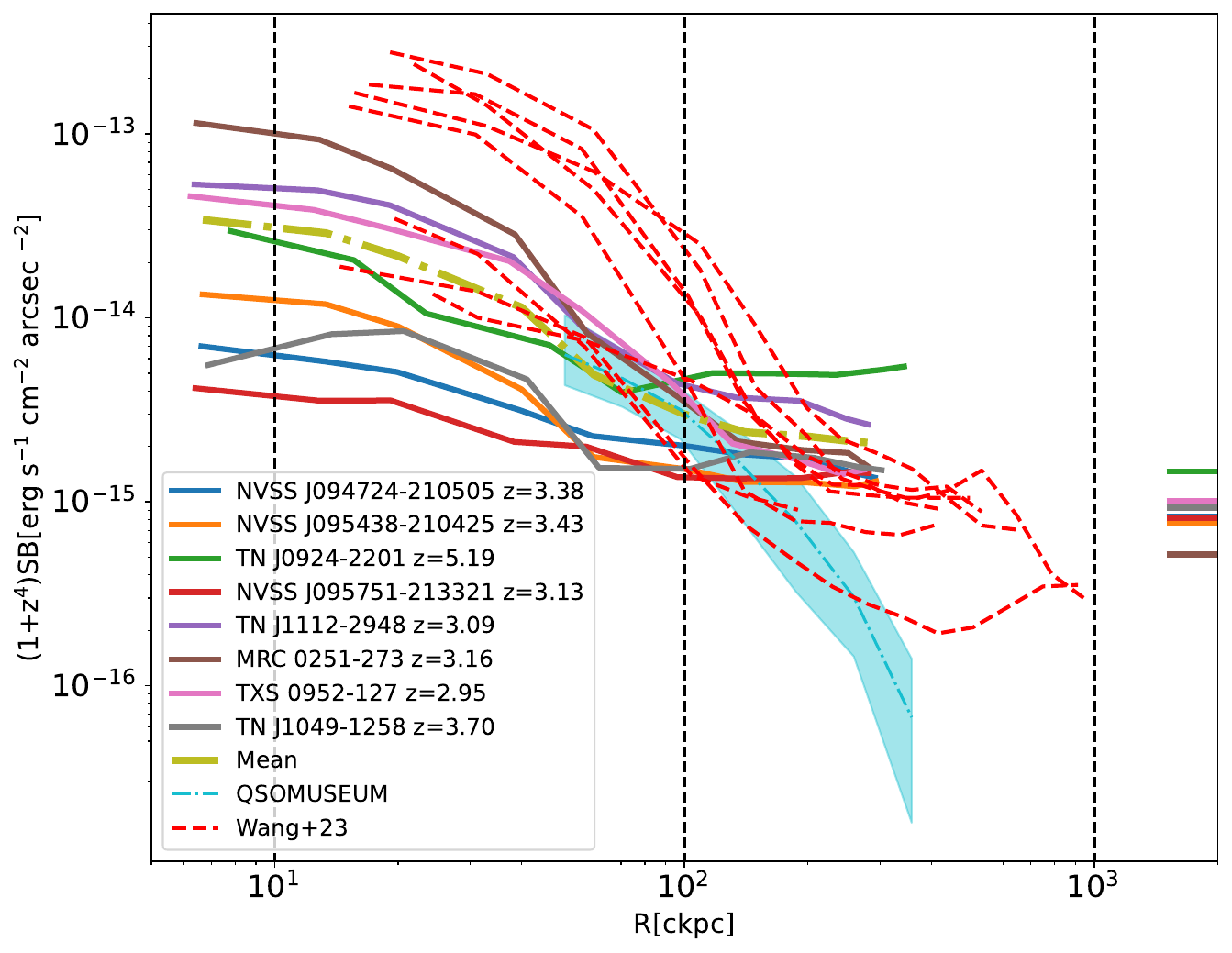}
    \caption{Redshift-corrected surface brightness curves for all the observed objects within our sample as well as those studied by \citet{Wang_2023} (dashed lines) and \citet{Arrigoni_Battaia_2018} (average shown as the solid blue area with dashed line). On the right side Y-axis we report the different background level estimated for each individual source.}
    \label{fig:allsamples}
\end{figure}

The collection of the SBPs obtained for all sources are shown in Fig.\ref{fig:allsamples}. In this Figure we also report the background levels in each image, shown as horizontal lines in the right vertical axis. We define the maximum nebular radius as the distance at which the measured surface brightness is twice the value of the background. Overall, the SBPs show a significant curvature in the log $SB$ - log $r$ representation, indicating that they are better reproduced by an exponential function rather than a power law. This is supported by the fact that the $\chi^{2}$ statistic is, on average, more than a factor of 10 larger when fitting with a power law. The SBPs are then fit with a function in the form $SB(r) = C$exp$(-r/r_{h}$) where $C$ is the normalization and $r_{h}$ is the scale length. The values of $r_{h}$ as well as the maximum measurable distance within the 2$\sigma$ limit of the nebulae are reported in Tab. \ref{tab:sourceprop}. In this Table we also give the derived total \lya\ luminosity as well as the maximum extent of the radio emission.

\begin{table}[ht]
    \caption{Properties of the brightness profiles.} 
    \centering
    \begin{tabular}{l| c c c c }
    Object & log L$_{{\rm Ly}\alpha}$ & r$_{h}$ & r$_{Max, Ly\alpha}$ & r$_{\rm Radio}$ \\
           & erg s$^{-1}$ & ckpc & ckpc & ckpc  \\
        \hline
NVSS~J151020-352803 & 43.35 &$^{a}$  &  57.0        & 58 \\
TXS~0952-217        & 43.63 & 33.6   &  21.3        & $^{d}$\\
TN~J1112-2948       & 43.47 & 31.8   &  28.1        & 83 \\
NVSS~J095751-213321 & 42.32 & \multicolumn{2}{c}{$^{b}$}               & 129 \\
MRC~0251-273        & 43.73 & 25.4   &  16.3        & $^{d}$\\
NVSS~J094724-210505 & 42.54 & 29.4   &  52.9        & 46 \\
NVSS~J095438-210425 & 42.77 & 16.7   &  15.1        & $^{d}$\\
TN~J1049-1258       & 42.98 & 28.3   &  10.1$^{c}$  & 92 \\
TN~J0924-2201       & 42.94 & \multicolumn{2}{c}{$^{b}$} &  10 \\
\hline
    \end{tabular}
    \tablefoot{Column description: logarithm of the total \lya\ luminosity in erg s$^{-1}$, scale length of the SBP, maximum extent of the \lya\ emission, and maximum size of the radio emission all in ckpc.\\
    \tablefoottext{a}{Due to the double lobe morphology of this source, the exponential fit radius was not derived.}
    \tablefoottext{b}{These are compact \lya\ sources and no estimate of their size can be derived.}
    \tablefoottext{c}{The size of the extended red emission region reported in \citealt{puga2023extended} is not included in this measurement, as its origin is yet to be clearly ascertained. As such, the size is limited to that traced by the host galaxy emission near the given redshift.}
    \tablefoottext{d}{Unresolved radio sources.}
    }
    \label{tab:sourceprop}
\end{table}

The closest comparison of our results can be drawn with the sample studied by \citet{Wang_2023}, given that they observe the same class of object (HzRGs) with the same instrument (MUSE). However, these objects were chosen for observations as part of the MUSE commissioning program due to their known brightness and extension. This induces a bias, which the authors themselves acknowledge, as they find their \lya\ nebulae to be both systematically brighter and larger than any other sample of high redshift AGN. Indeed, our targets' \lya\ emission is, as a general trend, significantly fainter: at a fixed radius of 50 ckpc we find values in the range $2 \times10^{-15} - 2\times 10^{-14}$ erg s$^{-1}$ cm$^{-2}$ arcsec$^{-2}$ while the observed SBPs of \citeauthor{Wang_2023} span a range a factor $\sim$10 brighter. Their SBPs are best reproduced by a two-piece function, including an exponential and power law. Given that our objects were fit exclusively with exponential functions, we limit the comparison to the results of their exponential fitting, where they find the scale lengths to vary in a range from 28 to 135 ckpc, averaging out at $\sim$75 ckpc. By comparison, our objects only span from 16 to 33 ckpc, and their average length is $\sim$27 ckpc, which is likely a slight overestimation of their average size given that two out of nine objects remain unresolved. Thus, these sources are associated with both significantly brighter and more extended \lya\ nebulae. The inner regions of the SBP of our HzRGs are closer to those seen in large samples of high redshift type 1 AGN \citep{Arrigoni_Battaia_2018}, though the objects in our sample and HzRGs more generally appear larger in size. The SBPs of both of these samples are also included in Figure \ref{fig:allsamples}

\subsection{Velocity and FWHM maps}
\label{vmaps}

Interpreting the kinematics of the \lya\ emitting gas is not, in general, a trivial effort. Resonant scattering can and regularly does result in spectra that are radically different from the intrinsic \lya\ emission of the galaxy due to the presence of intervening HI gas at both extra- and intergalactic scales. In order to obtain robust and representative results of the kinematic properties of the ionized gas, we chose to forgo some amount of spatial resolution in order to increase the signal-to-noise of the spectra used in our analysis. To this end, we performed Voronoi tessellation \citep{Cappellari2003} of the observed emission by first making a hard cutoff of S/N$>$1.5 to limit the pixels used in the tesselation, then adjusting the desired S/N in each bin in order to achieve the best compromise between the spatial resolution and quality of the spectra.

Once the bins have been selected, we perform Gaussian fitting of the different spectra. In those cases where the total S/N is high enough, it is possible to account for absorption troughs and multiple emission components. The results of this analysis are presented in velocity and line width maps in Appendix A. In the case of NVSS J095751-213321, although the emission is visibly extended, it is faint enough that creating more than a single bin resulted in excessively noisy spectra which were not fit for modeling. Regarding TN J0924-2201, the object is not spatially resolved. In both of these cases, only the width of the line is given as a representative value of the source in \hyperref[appendixB]{Appendix B}.

The results of this analysis are \lya\ line widths which oscillate in the range between 900 and 1,400 \kms\ with relatively small changes across individual nebulae. The velocity fields have generally very low amplitudes, typically smaller than $\pm$100 \kms. In three  cases out of the six for which this analysis is possible, namely TN~J1112-2948, TN~J1049-1258, and MRC~0251-273, there are clear signs of ordered rotation, albeit maintaining the aforementioned small-scale gradients.

It is also interesting to explore whether there is any connection between the gas kinematic and the radio emission, i.e., whether there are signs of interaction between the radio jets and the ionized gas. We are able to simultaneously resolve the radio and velocity fields only in three cases, and none of them display a clear correlation between the radio morphology and the morpho-kinematic properties of the host galaxy, be it systemic velocity or line width. We describe below this comparison in more detail:

\begin{itemize}
    \item \textbf{NVSS J151020-352803}: The radio emission covers almost the entirety of the observable nebula. The western lobe is redshifted by some $\sim$200\kms with respect to the eastern lobe. It also appears to be more disturbed, but, as absorption correction was not possible in this case, it is difficult to draw any conclusions.
    \item \textbf{TN J1112-2948}: In this case, the radio lobes are well beyond the observed nebula. A clear rotational motion is observed in the velocity field of this object. Regarding the FWHM, it does appear to have a slight gradient along the direction of the radio jet.
    \item \textbf{TN J1049-1258}: The kinematics and general characteristics of this peculiar case were studied more in depth in \citet{puga2023extended}. In summary, the host galaxy displays a clear rotational motion with a wide ($\sim$10$^{3}$\kms) line profile, while the extended emission has a velocity gradient along the radio axis and a much narrower ($\sim$300\kms) profile.
\end{itemize}

There are three objects that did not contain enough flux in the entirety of the nebula to produce more than one bin in the cases of NVSS J095751-213321 and TN J0924-2201, or two bins in the case of NVSS J095438-210425. In these instances, we forgo the spatial kinematic analysis, as a single bin contains no spatial information while the two bins in NVSS J095438-210425 are very spatially extended and contain many loose pixels. Thus, they are highly likely to smother out any kinematic features along the breadth of the nebula.

\begin{table*}[ht]
\caption{Emission line fluxes and ratios}
\centering
    \begin{tabular}{l | r r r r r r r r}
Object & F$_{\rm Ly\alpha}$ & F$_{\rm C~IV}$ & F$_{\rm He~II}$ & F$_{\rm C~III}$ & \lya/CIV & \lya/HeII & CIV/HeII & CIII]/CIV \\\hline
NVSS~J151020-352803 & 317$\pm$11 & 45$\pm$10 & $\leq$11 & ---       & 7   & $\geq$28 & $\geq$4 & ---\\
TXS~0952-217        & 770$\pm$6  & 70$\pm$4  & 62$\pm$5 & 47$\pm$6  & 11  & 12      & 1.1      & 0.75 \\
TN~J1112-2948       & 430$\pm$10 & 71$\pm$6  & 65$\pm$6 & 55$\pm$20 & 6   & 7       & 1.1      & 0.84 \\
NVSS~J095751-213321 & 15$\pm$2   &  ---      &  ---     & ---       & --- & ---     & ---      & --- \\
MRC~0251-273        & 790$\pm$12 & 117$\pm$4 & 49$\pm$5 & 32$\pm$10 & 7   & 16      & 2.4      & 0.65\\
NVSS~J094724-210505 & 32$\pm$4   & ---       & ---      & ---       & --- & ---     & ---      & --- \\
NVSS~J095438-210425 & 53$\pm$3   & 26$\pm$3  & $\leq$6  & ---       & 2   & $\geq$9 & $\geq$4  & --- \\
TN~J1049-1258       & 73$\pm$4   & ---       & ---      & ---       & --- & ---     & ---      & --- \\
TN~J0924-2201       & 30$\pm$3   & ---       & ---      & ---       & --- & ---     & ---      & --- \\
\hline
\end{tabular}
\tablefoot{List of the emission line fluxes (in units of 10$^{-18}$ erg cm$^{-2}$ s$^{-1}$) measured within the 2$\sigma$ region of emission for \lya.}
\label{tab2}
\end{table*}

\subsection{Emission line fluxes}
\label{lineratios}

The lines available in the MUSE spectra of these objects are all at rest-frame UV, namely, \lya, NV$\lambda\lambda$1238,1242, CIV$\lambda\lambda$1548,1550, HeII$\lambda$1640, and CIII]$\lambda$1908. The emission line fluxes are measured for each object by adding the spectrum in each pixel where the \lya\ flux remains above 2$\sigma$. This ensures that the signal-to-noise ratio is optimized for this line. Regarding the other emission lines, which are visibly more compact, it guarantees all of their emission will be taken into account, even if it increases the final uncertainty. We found this to be a desirable compromise: resonant scattering means precise and complex modeling is necessary for \lya, while a simple Gaussian model will quite accurately reproduce the observed profile of the other lines and will not be as affected by pixels which include spurious noise or whose flux is close to background levels.

For those objects which do not display visible absorption troughs, either because they are too faint or too heavily absorbed in one side, a single Gaussian is used to fit \lya. If the line appears absorbed but is too faint for accurate fitting, a narrowband measurement is used.

There is one exception to this rule, which occurs on the brightest source of this fainter population, TN 1049-1258. In this object, we found He~II emission, which was not associated to the host galaxy but rather to an extended region of heavily red-shifted and extended \lya\ emission, aligned with the radio emission. Given that the origin and nature of this emission are still under discussion, we refer the reader to \citealt{puga2023extended}.

\subsubsection{Line ratios and ionization mechanisms}

Out of nine sources in our sample, five display detectable levels of line emission besides \lya (see Table \ref{tab2}). The observed \lya/CIV and \lya/HeII ratios are generally within the observed range for other HzRGs. For instance, \citet{Villar_Mart_n_2007_a} find mean values of \lya/CIV=7.8 and \lya/HeII=9.6 in a sample of 53 HzRGs at z$\sim$2-4. We find mean values of \lya/CIV=6.6 and \lya/HeII=14.4, albeit from a much more limited sample. In the cases of TXS 0952-217 and MRC 0251-273, the relatively high \lya/HeII and \lya/CIV ratios suggest an alternative origin for at least a fraction of the \lya\ emission, meaning star formation is a possible additional ionization mechanism as postulated by \citet{Villar_Mart_n_2007_a}. 

Given the \lya/HeII and \lya/CIV ratios observed in the brighter objects of our sample (with the exception of NVSS J151020), and those observed in large samples of HzRGs, this is expected. If one assumes relatively similar ratios for all sources, the emission of the non-\lya\ lines on the fainter part of the sample will fall to or below background levels, rendering them unobservable. This presents an additional issue: when observing this fainter end of the HzRG \lya\ luminosity function, we become limited to highly uncertain upper bounds for the fluxes of these lines. Consequently, whether or not the ionization mechanism for \lya\ is the same as for the more luminous sources remains a point of contention as precise diagnostics become unfeasible. It could very well be that in these fainter objects the line ratios are much higher than in those on the brighter end and that the sources of ionizing photons are altogether different; the available observations cannot shed light on this.

\citet{Matsuoka_2009} proposed a combination of the CIV/HeII and CIII]/CIV line ratios as a probe for metallicity in high redshift active galaxies. While the estimated metallicities are dependent on the parameters used in the photoionization models, we find our observed line ratios to be well in line with those in the sample presented in \citet{Matsuoka_2018}.

\subsubsection{The CIV/CIII] ratio and UV photon escape fraction}

Recent works, such as those by \citet{Schaerer_2022} and \citet{Saxena_2022}, have found a likely correlation between the strength of the CIV$\lambda\lambda$1548,1550 emission line and the escape fraction of Lyman continuum, both at low and high redshift. MUSE observations are uniquely suited as high signal-to-noise observations allow for the complete modeling of the \lya\ profile. Consequently, one can directly obtain f$_{esc}$ by comparing the intrinsic and observed fluxes of the line, and measure the strength of the CIV line by comparing it with CIII]$\lambda$1908. \citealt{Schaerer_2022} proposes an empirical cutoff of C43=CIV/CIII]>0.75, above which galaxies become strong Lyman leakers (f$_{esc}$>0.1).

In our case, an accurate and complete modeling of the \lya\ emission was only possible for two sources; MRC 0251-273 and TXS 0952-217. In both cases, C43 is above the cutoff, with values of 3.31 and 1.25 respectively, while the escape fractions of \lya\ photons are 0.59 and 0.41

These estimates must be taken with a grain of salt, as, in addition to the intrinsic uncertainties that come from working with such faint objects, the results of the \lya modeling are affected by strong degeneracies between intrinsic emission and HI column density (e.g. \cite{Silva_2017}). As such, they are best interpreted as upper bounds of the true value of \fesc.

\section{Radio versus \texorpdfstring{\lya}{} properties}
\label{radiovslya}
\subsection{Radio versus \texorpdfstring{\lya}{} luminosities}

\begin{figure}[tp]
    \centering
    \includegraphics[width=0.45\textwidth]{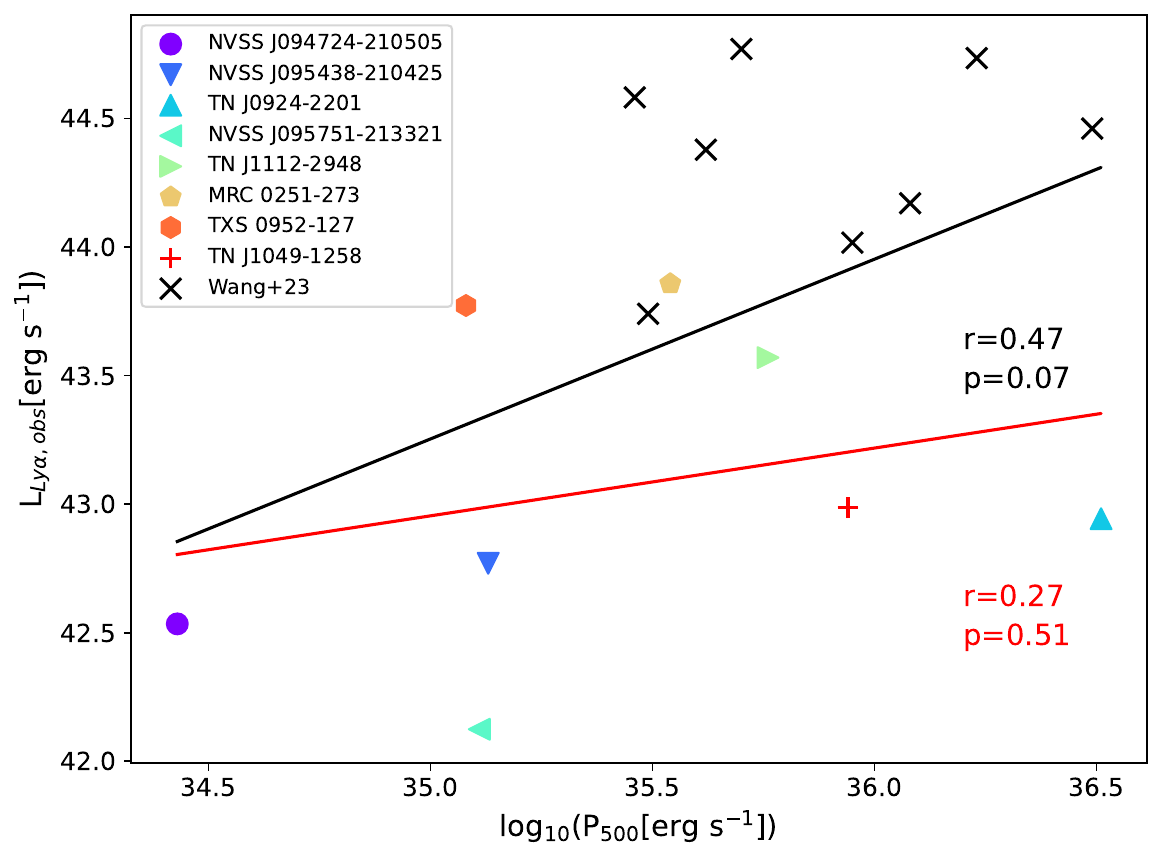}
    \caption{Total \lya\ luminosity vs radio power at 500 MHz. We added to the measurements of our sample (colored symbols) those from \citet{Wang_2023} marked as black crosses. The red line is the best linear fit to our data only, while the black line displays the fit to the combined sample.}
    \label{fig:radvslya}
\end{figure}

Low redshift radio galaxies show a strong trend of increasing optical line luminosity with radio power (see, e.g., \citealt{baum89a,baum89b,rawlings89,willott99}). \citet{buttiglione10} explored this connection based on a spectroscopic study of the radio sources in the Third Cambridge Catalog (3C, \citealt{spinrad85}) with z$<$0.3. They found that radio galaxies follow a quasi-linear correlation between line ([O III]$\lambda$5007 and H$\alpha$) and radio luminosity at 178 MHz. i.e., $L_{\rm [O~III]} \propto L_{178}^{0.99}$ with a spread of 0.5 dex.
\citet{capetti23} found from observations of 3C radio galaxies at 0.3$<$z$<$0.8 that although the trend of increasing optical emission line (also [O III]$\lambda$5007) luminosity with radio power is still present, the slope of radio-line correlation decreases at the highest radio power, $\sim 10^{28} \WHz$. This suggests that in more luminous AGNs a higher fraction of nuclear ionizing photons escape without being absorbed and reprocessed by the interstellar medium. The study of HzRGs enables us to explore this issue at even higher radio power. 

In Fig. \ref{fig:radvslya} we compare the radio power at the rest frame frequency of 500 MHz as estimated by \citet{Miley_2008} and the total \lya\ luminosity. We also include in this comparison the measurements by \citet{Wang_2023}. By considering only the nine sources observed within our program, we found no connection between these quantities. The inclusion of the eight sources from \citet{Wang_2023} drives a marginal correlation with a 5\% probability and a slope of m=0.75$\pm$0.33.

The spread of the data points is very large, about two orders of magnitude, which is much larger than what is observed in lower redshift RGs. This indicates a looser connection between the jet power and the \lya\ line luminosity and might be due to either 1) the presence of additional ionization mechanisms (e.g. star formation) or 2) internal absorption of the \lya\ photon within the gas nebulae, due to either the aforementioned resonant scattering or absorption by dust. This second mechanism could explain the systematic spread of these data, as variations in the geometry and density of the HI gas in both the ISM and CGM (\citealt{Verhamme_2006}, \citealt{Behrens_2014}, \citealt{Gronke_2015}, \citealt{Gronke_2017}), as well as galaxy-to-galaxy differences in the concentration of dust (see \citealt{Laursen_2009}) will yield varying escape fractions of \lya, while more "traditional" ionized emission lines such as [O III]$\lambda$5007 and H$\alpha$ are not subject to those effects to such a high degree.

\subsection{Geometrical relation between radio and \texorpdfstring{\lya}{} emission}

\citet{Wang_2023} report of a connection between radio and \lya\ emission, with an alignment between PA of the radio emission and of the \lya\ nebula. In our sample of nine HzRGs there are six sources showing a resolved structure in the VLASS images, namely NVSS~J151020-352803, TN~J1112-2948, NVSS~J095751-213321, NVSS~J094724-210505, TN~J1049-1258, and TN~J0924-2201.\footnote{In the VLASS image this source is unresolved but a higher resolution image (FWHM = 0\farcs49) at 8.5 GHz available at https://www.vla.nrao.edu/cgi-bin/nvas-pos.pl shows two components separated by 1\farcs2, oriented at PA$\sim 70^\circ$.} However, in J0924-2201, the farthest source of the sample, the \lya\ line is extremely compact and very faint, meaning no information can be extracted about the morphology of its \lya\ emission, so it is dropped from this analysis.

In order to estimate the angle of the \lya\ nebulae, we perform a Gaussian 2D fitting using the narrowband images shown in the Appendix A (which are smoothed using a 1-pixel Gaussian kernel), yielding both the PA as well as the semimajor and semiminor axes of the ellipse. In some cases, the available signal and the small size of the nebula do not lead to a robust fit. In those instances, the PA is extracted from the morphology of the 2$\sigma$ isophote. Finally, although in NVSS~J095751-213321 the \lya\ emission is slightly extended, it is far too faint for any robust analysis (see Fig. A.4). The derived values are listed in Table \ref{tab:radiovslya}.

In the radio, all the extended sources present a clear double-lobe morphology and the PA was calculated using the brightest pixel of each lobe.

\begin{table*}[ht]
\caption{Geometrical properties of the \lya\ nebulae and radio emission of the sources for which both are resolved.}
\centering
    \begin{tabular}{l|c|c|c|c|c}
        Object & $\phi_{Ly\alpha}$ & $\phi_{Radio}$ & |$\phi_{Ly\alpha}-\phi_{Radio}$| & $\alpha_{Ly\alpha}$ & R$_{Max, Ly\alpha}$/R$_{Max, Radio}$\\
        \hline
        NVSS J151020-352803 & 19 & 15 & 4 & 0.78 & 0.97 \\
        TN J1112-2948 & 13 & 27 & 14 & 0.92 & 0.33 \\
        NVSS J094724-210505 & 17 & 19 & 2 & 0.45 & 1.15\\
        TN J1049-1258$^{a}$ & 0 & 16 & 16 & 0.39 & 0.1\\
 \hline
 \end{tabular}
    \tablefoot{Column description: name, PA of the \lya\ nebula, PA of the radio emission, difference between the two PAs, ratio between the extent of the \lya\ and radio emission.\\
    \tablefoottext{a}{In this case, the extended redshifted emission was used, as described in \citet{puga2023extended}.}
    }
    \label{tab:radiovslya}
\end{table*}

\begin{figure*}[ht]
    \centering
    \includegraphics[width=0.42\textwidth]{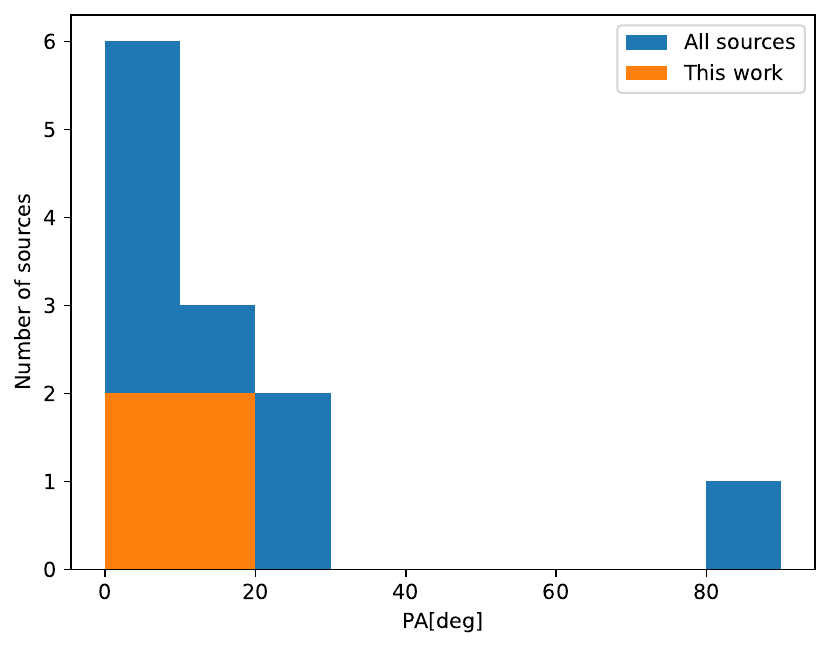}
    \includegraphics[width=0.52\textwidth]{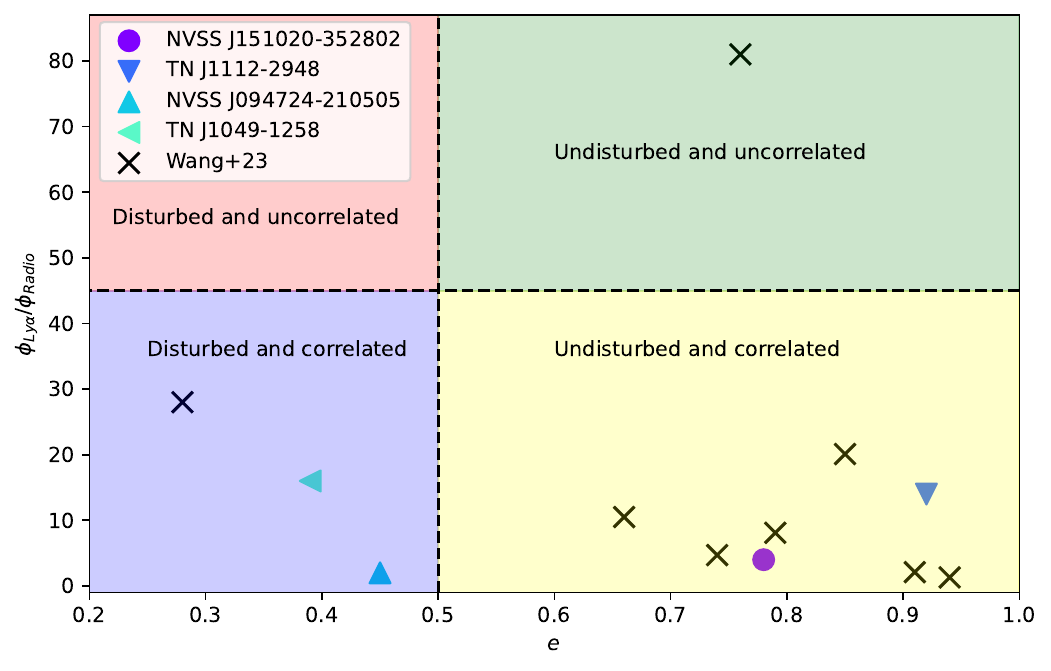}
    \caption{Left: offset between the PA of the radio and of the \lya\ structures. The orange histogram refers to the sources of our sample, the blue histogram consider also those from \citet{Wang_2023}. Right: nebular ellipticity $e$ vs PA offset.}
    \label{fig:pa-radiovslya}
\end{figure*}

All four sources in our observing program where we are able to derive this comparison show an alignment between these two structures within less than 16$^\circ$. We also consider the measurements by \citeauthor{Wang_2023} and in Fig. \ref{fig:pa-radiovslya} we show the distribution of the offsets of PAs between the radio axis and of the \lya\ nebulae. We performed a Kolmogorov-Smirnov statistical test on the distribution of PA offsets for both the total sample (our work plus W23+) and W23+ alone, as it is the larger individual sample. We report the probability of drawing such a distribution from a random sample as P=0.7\% for the total sample and P=10\% for W23+ alone. As such, the four sources added by our analysis are key in increasing the statistical significance for the observed alignment between radio and \lya\ emission.

From the point of view of size, there is more variety. In some cases, such as NVSS~J151020-352803 and NVSS~J094724-210505, the expanse of their radio and ionized emissions are very similar. The other two, however, present radio emission much larger than the observed \lya. The objects which did not form part of this size analysis must also be accounted for, as there are multiple sources in which the \lya\ emission is extended while the radio emission remains unresolved. Therefore, we cannot claim that there exists a connection between the extension of the ionized gas emitting regions and the radio emission, but the evidence is strong that alignment between radio and \lya\ is present in HzRGs. 

To analyze the relation between the alignment of radio and \lya\ emission with respect to the general nebular morphology we use the ellipticity from the 2D Gaussian analysis as a measure of the asymmetry of the emission. When comparing this alignment parameter, we observe clustering in a region which corresponds to nebulae whose emission is highly correlated to the radio in terms of direction, but remain relatively symmetrical nonetheless. While the radio emission and its associated phenomena are likely to have an enhancing effect on the emission of the ionized gas they interact with, they do not appear to be reshaping the nebulae on larger scales; at the very least, this is not the case in the majority of sources. There are exceptions to this: three sources are highly elongated along the direction of the radio emission, and a single source appears to be oriented perpendicular to the radio.

These results, and in particular those drawn from our sample given the lower exposure times and intrinsic brightness of the sources, must be taken with a caveat: it is exceedingly difficult to draw robust limits on the extent of these nebulae, and radiative transfer of \lya\ plays a considerable part in reshaping the original ionized emission, in both kinematic and morphological terms. This is compounded by the fact that there are many phenomena associated with radio jets which could be enhancing ionized line emission, and, in fact, it is likely they all play a part: relativistic beaming, cavities and their associated shocks, and young stellar emission from positive feedback, among others. Disentangling the extent to which these components contribute to this apparent alignment is beyond the scope of this work and will require multi-wavelength observations.

\citet{Villar_Martin_2007_b} observed similar phenomena occurring in a smaller sample of HzRGs (z$\sim$2.5), where two of the three nebulae observed with VIMOS were well aligned with the radio direction. In one object the \lya\ emission remained relatively symmetrical, while in a different one it was highly elongated along the radio axis.

Alignment between the morphologies of the extended line emission and the radio emission around high redshift active galaxies is a well documented phenomenon (e.g. \citealt{Heckman_1991,Nesvadba_2017}); \citet{Eales_1992} proposed the origin for these observations to be partly due to selection effects. Radio emission appears to be preferentially observed when aligned with nebular line emission due to jet-cloud interactions in high-density environments. These denser regions, which are inherently brighter due to both interaction with the jets and illumination from the AGN, enhance the radio emission itself, leading to an inherent bias born from observational limitations; as we probe the brighter end of the \lya\ luminosity function of HzRGs, we are set to observe these objects in which this radio emission is enhanced. This is in agreement with the fact that no clear correlation has been observed between the sizes of the radio and nebular emissions. If the origin for the nebular morphology were jet-cloud interactions, one would also expect a spatial correlation.

It is highly likely that the different environments, AGN activity cycles and their associated feedback, and accretion histories of these galaxies all play a part in the observed nebular morphologies. Disentangling the various underlying components requires a very profound multi-wavelength approach and is beyond the scope of this work.

\section{Summary and conclusions}
\label{summary}

In this study, we analyze the IFU data of 11 observations of HzRGs performed with MUSE at the VLT. We detect 9 sources and provide upper limits to the emission of two undetected objects for which we cannot confirm the identification as HzRGs. This paper covers the \lya\ morphology and spatial distribution, the overall UV spectral properties of the sources and the possible connections of these parameters with the radio emission associated with these objects. Our summarized findings are as follows:

\begin{itemize}
\item We present synthetic narrowband \lya\ maps, nuclear spectra and surface brightness profiles for all the observed sources in \hyperref[appendixA]{Appendix A}. For those sources in which HeII is detected, we report accurate systemic redshift measurements.
\item We find the SBPs of the objects in our sample are generally better described by an exponential function. They are systematically smaller and fainter than those previously studied (e.g. W23+). These latter sources, however, were pre-selected for having bright \lya\ nebulae of large extension, and we conclude this is a result of selection bias. The SBP of our sources are instead similar to those observed in type 1 QSOs (e.g. \citealt{Arrigoni_Battaia_2018}). 
\item The general spectral properties of the sample are in line with other observations of \lya\ in HzRGs, presenting clear signs of absorption due to intervening HI gas and very wide line profiles in the order of 1,000 \kms. In terms of velocity fields, we observe rotation of very small amplitude in a number of objects and less ordered kinematics in others, without a clear connection to radio power and morphology or \lya\ luminosity.
\item The observed emission line ratios are, in most cases, in agreement with an AGN as the source of ionizing photons. We find slightly increased \lya/HeII ratios in a number of objects, but these also coincide with an increase in the CIV/HeII ratio while \lya/CIV remains within levels expected from AGN photoionization. As such, the ratios can be explained through either a higher ionization parameter as the result of, for instance, increased star forming activity \citep{Villar_Mart_n_2007_a}, or an overall lower metallicity \citep{Matsuoka_2009}.
\item Contrary to what is found in lower redshift sources using rest-frame optical ionized lines, we do not report a strong correlation between the total or nuclear \lya\ luminosities and radio power. This is possibly due to the effects of resonant scattering of \lya\ photons, yielding luminosities which are highly dependent on both the intervening HI gas along the line of sight, the environment and morphology of the nebulae of origin and the presence of dust in the host galaxy.
\item For the four objects in our sample in which it was possible to simultaneously spatially resolve the radio and nebular UV emission, we find all of them to be very well aligned (with, at most, a 16$^\circ$ deviation) with the direction of the radio emission. This is a well documented and commonly observed phenomenon which has been postulated as an effect of limiting magnitudes and selection bias due to jet-cloud interactions. While we probe a slightly lower power sample than has previously been done, this effects remains pervasive, as we also fail to observe a clear correlation between the degree of alignment and the morphological properties of the nebulae.
\end{itemize}

In general, we find that while these data allow for, in particular cases, the detection and characterization of AGN-related phenomena (e.g. \citealt{Silva_2017,puga202420}), these are far from the norm. Most of the objects in this sample are far too faint to allow for such an analysis given the observing strategy, which was more "snapshot-minded" than was needed for both attaining a high enough signal near the nuclear regions in order to confirm the presence of outflows and quantify their effects, as well as probing the very sparse and extended emission line regions observed in other samples of HzRGs. The advent of new generation telescopes, JWST and ELT in particular, should allow for more accurate IFU observations and yield better estimates of the impact of AGN feedback at this essential point in cosmic time.

One positive result is the confirmation that the alignment effect between radio and \lya\ keeps, even when probing the lower end of the available luminosity range for HzRGs. While this is not of interest to IFU and imaging projects, long slit spectroscopy can still play a role in identifying and characterizing these objects, as the radio morphology (and to a lesser degree, the radio power) remains a good primer for designing observing strategies which orient the slit along the radio axis. 

\begin{acknowledgements}

We extend our gratitude to Wuji Wang and Fabrizio Arrigoni-Battaia for allowing us to use their data in parts of our analysis, as well as to the referee for their help and insightful comments. The Pan-STARRS1 Surveys (PS1) and the PS1 public science archive have been made possible through contributions by the Institute for Astronomy, the University of Hawaii, the Pan-STARRS Project Office, the Max-Planck Society and its participating institutes, the Max Planck Institute for Astronomy, Heidelberg and the Max Planck Institute for Extraterrestrial Physics, Garching, The Johns Hopkins University, Durham University, the University of Edinburgh, the Queen's University Belfast, the Harvard-Smithsonian Center for Astrophysics, the Las Cumbres Observatory Global Telescope Network Incorporated, the National Central University of Taiwan, the Space Telescope Science Institute, the National Aeronautics and Space Administration under Grant No. NNX08AR22G issued through the Planetary Science Division of the NASA Science Mission Directorate, the National Science Foundation Grant No. AST–1238877, the University of Maryland, Eotvos Lorand University (ELTE), the Los Alamos National Laboratory, and the Gordon and Betty Moore Foundation. The Digitized Sky Survey was produced at the Space Telescope Science Institute under U.S. Government grant NAG W–2166. The images of these surveys are based on photographic data obtained using the Oschin Schmidt Telescope on Palomar Mountain and the UK Schmidt Telescope. The plates were processed into the present compressed digital form with the permission of these institutions.

\end{acknowledgements}

\bibliographystyle{aa} 
\bibliography{my2} %

\begin{appendix}
\section*{Appendix A: main analysis results}\label{appendixA}

\renewcommand{\thefigure}{A.\arabic{figure}}
\setcounter{figure}{0}

\begin{minipage}{1.0\textwidth}
  \strut\newline
  \centering
  \includegraphics[width=0.8\textwidth]{J151nuc.pdf}
  \includegraphics[width=.4\textwidth]{J151map.pdf}
  \hspace{1cm}
  \includegraphics[width=.5\textwidth]{J151brightdiff.pdf}\\
  \includegraphics[width=.9\textwidth]{J151vorvelmaps.pdf}
  
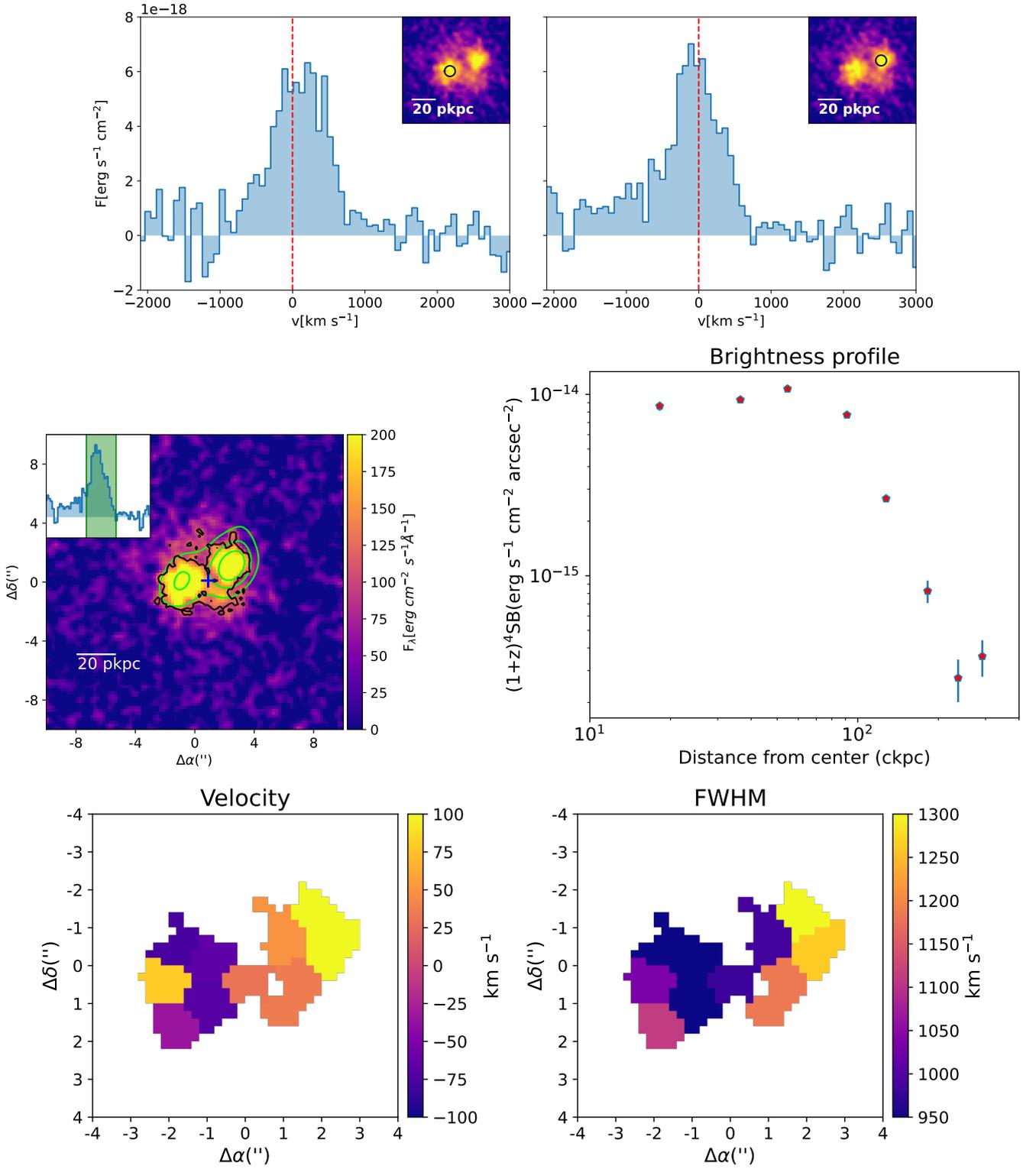
\captionof{figure}{NVSS J151020-352803. Top: spectra extracted from the two \lya\ regions (shown as the black circles in the narrow band images in the 10"$\times$10" insets). Middle: on the left is shown the \lya\ image extracted from the spectral region shown in the inset with superposed radio contours. The black contours mark the 2$\sigma$ flux limits, and the blue cross is the point from which the brightness profile is extracted. On the right is shown the surface brightness profile. Bottom: Voronoi-tessellated maps of velocity and FWHM of the \lya\ line.}\label{fig:j151nuc}
\end{minipage}

\begin{figure*}
    \centering
    \includegraphics[width=\textwidth]{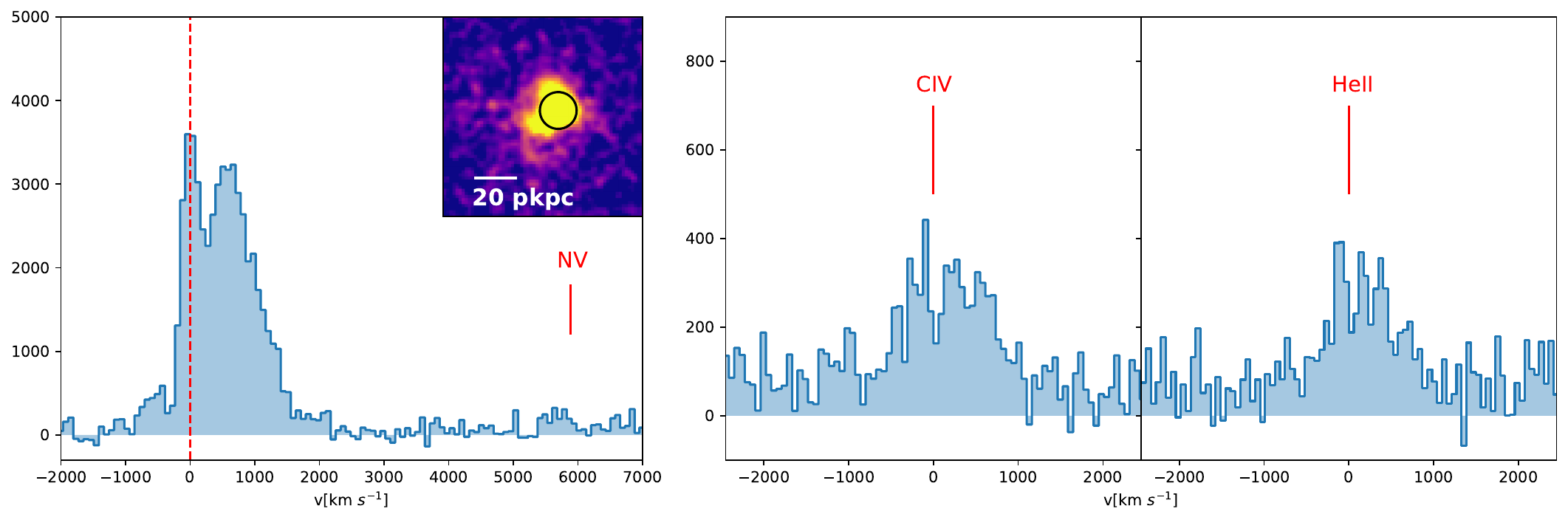}
    \includegraphics[width=.5\textwidth]{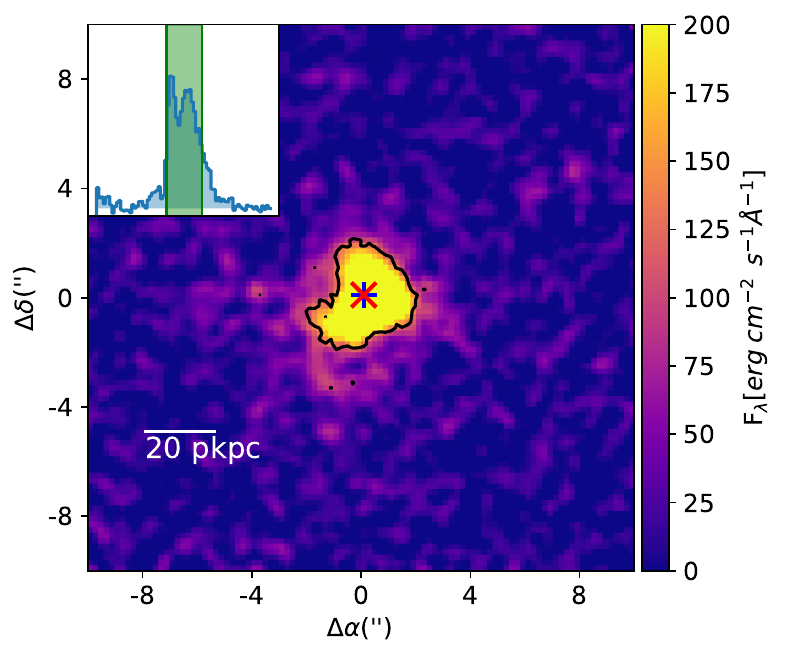}
    \includegraphics[width=.45\textwidth]{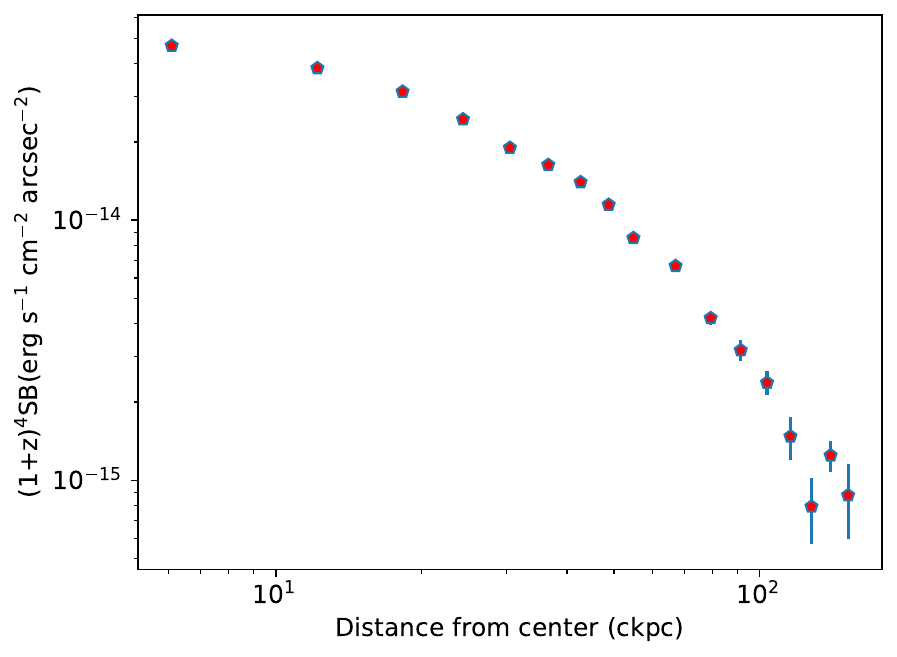}\\
    \includegraphics[width=\textwidth]{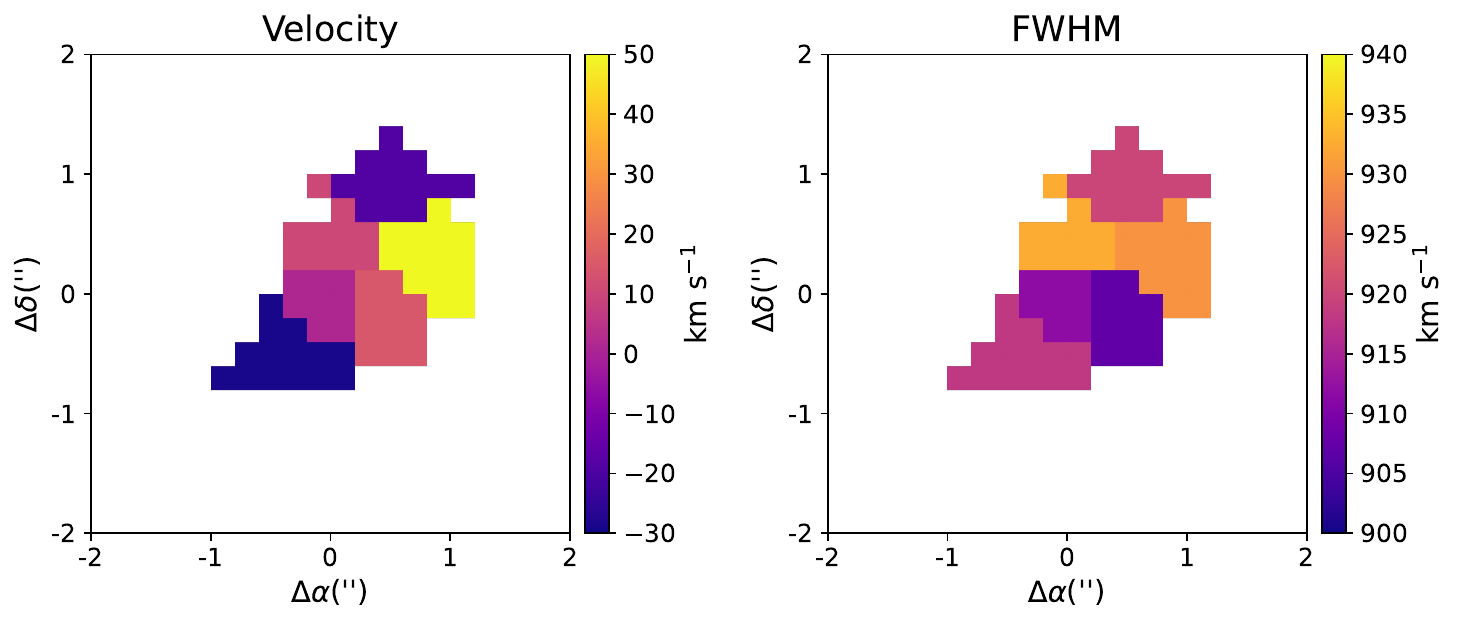}
    \caption{TXS~0952-217. Top: nuclear spectra centered on the \lya\ (left) and CIV + HeII lines (right) extracted from the region shown as black circle in the inset. Middle left: \lya\ image extracted from the spectral region shown in the 10"$\times$10" inset. The black contours mark the 2$\sigma$ flux limits, and the blue cross is the point from which the brightness profile is extracted. The location of the radio source, unresolved in the VLASS image, is marked with a red cross. Middle right: \lya\ brightness profile. Bottom: Voronoi-tessellated maps of velocity and FWHM of the \lya\ line.}
    \label{fig:txs095}
\end{figure*}

\begin{figure*}
    \centering
    \includegraphics[width=0.85\textwidth]{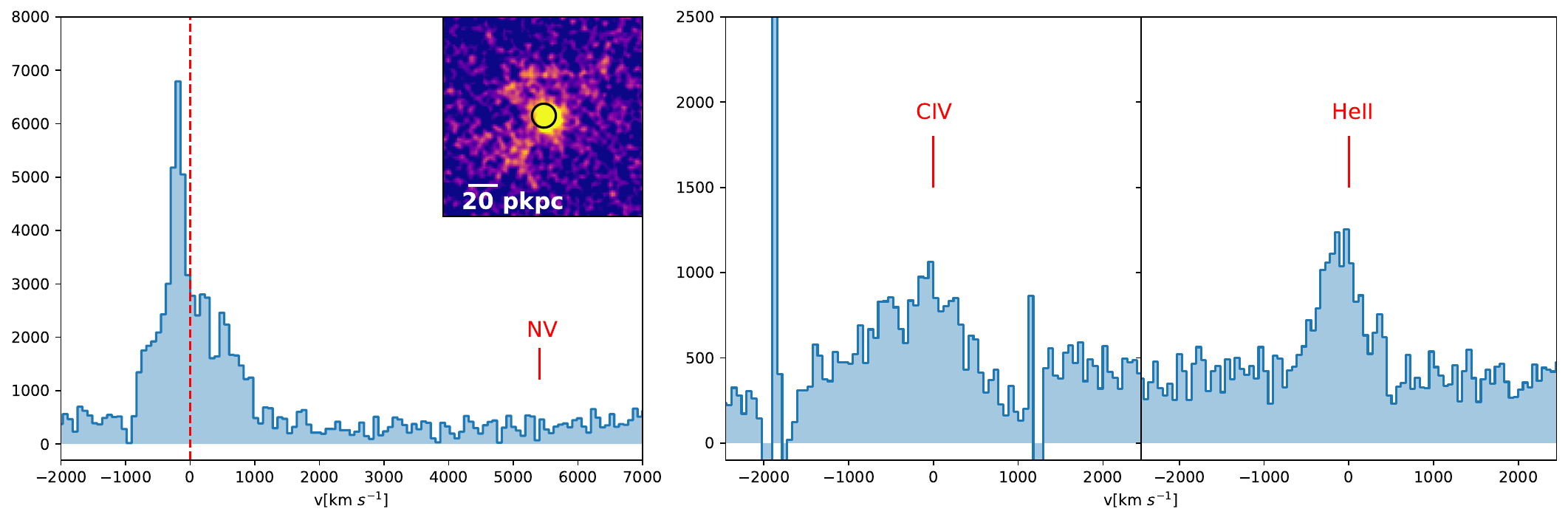}
    \includegraphics[width=0.85\textwidth]{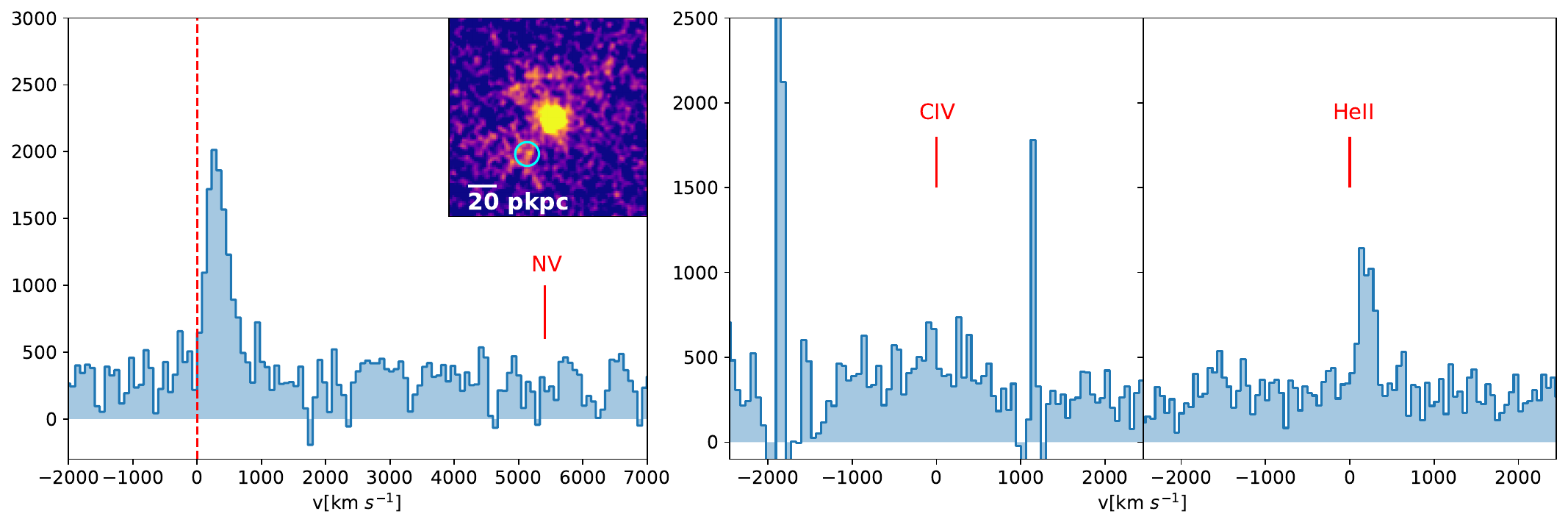} 
    \includegraphics[width=.37\textwidth]{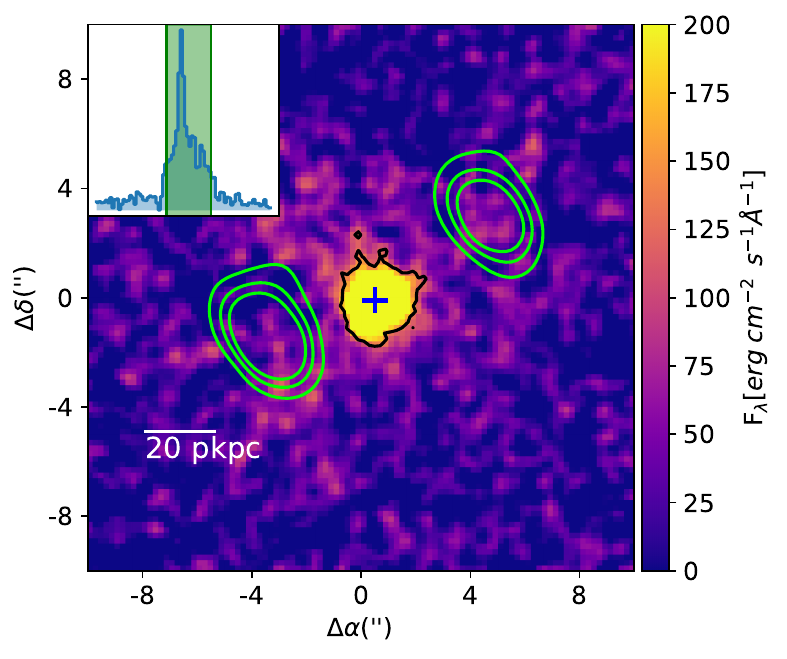}
    \includegraphics[width=.42\textwidth]{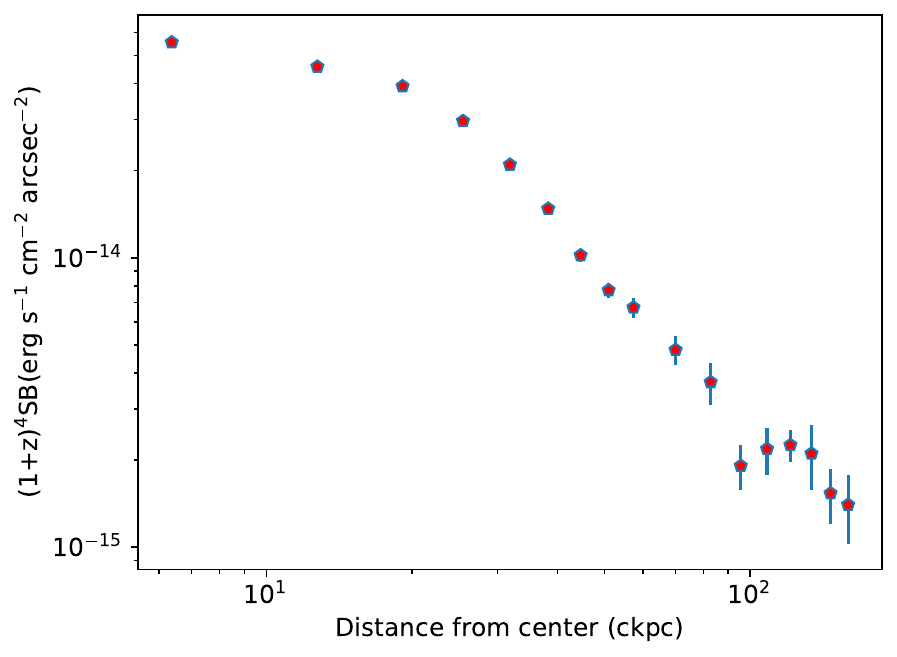}
        \label{fig:j111nuc}
    \includegraphics[width=0.85\textwidth]{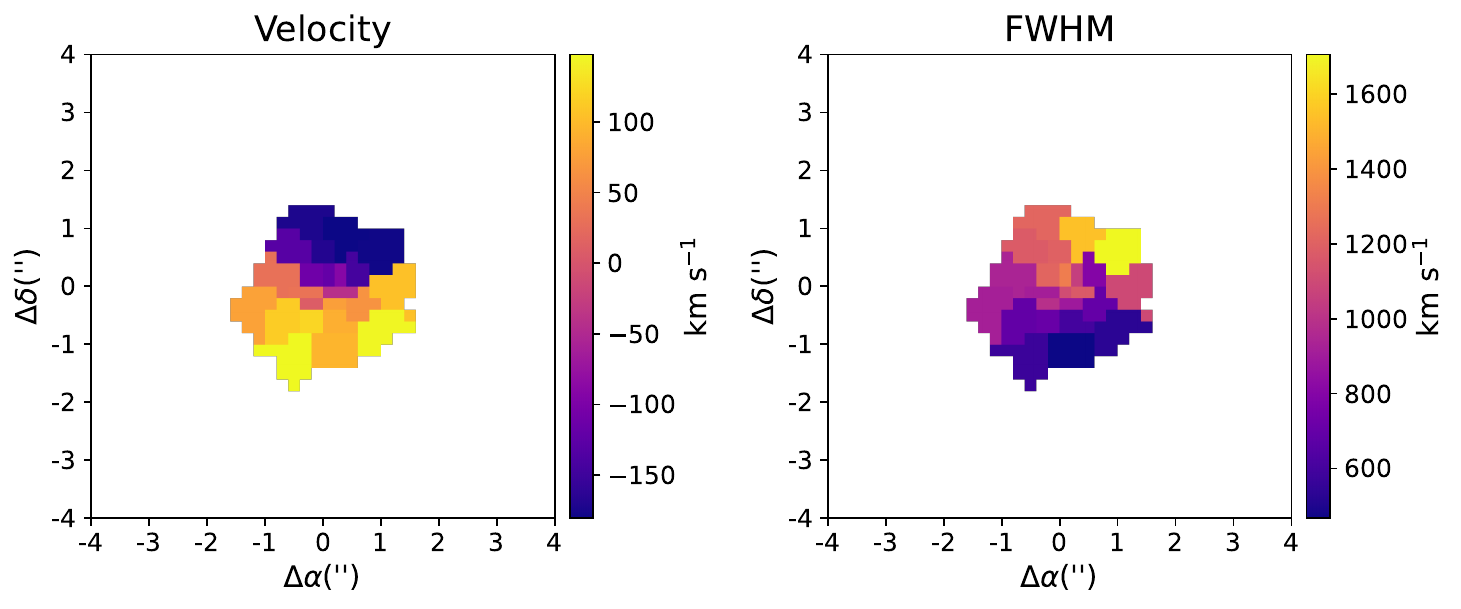}
       \caption{TN~J1112-2948. First line, left: nuclear spectra in the \lya\ spectral region extracted from a nuclear aperture of 1\farcs2\ radius (shown as the black circle in the narrow band image in the 20"$\times$20" inset) with the velocity scale based on the redshift measured from the He~II line. On the right, nuclear spectrum in the C~IV and He~II spectral region. Second line: same as the first for the companion galaxy. Third line, left: narrow band image extracted from the spectral range shown in the inset centered on the \lya\ line with superposed radio contours. The black contours mark the 2$\sigma$ flux limits, and the blue cross is the point from which the brightness profile is extracted. On the right, \lya\ brightness profile. Fourth line: Voronoi-tessellated maps of velocity and FWHM of the \lya\ line.}
\end{figure*}

\begin{figure*}
    \centering
    \includegraphics[width=.52\textwidth]{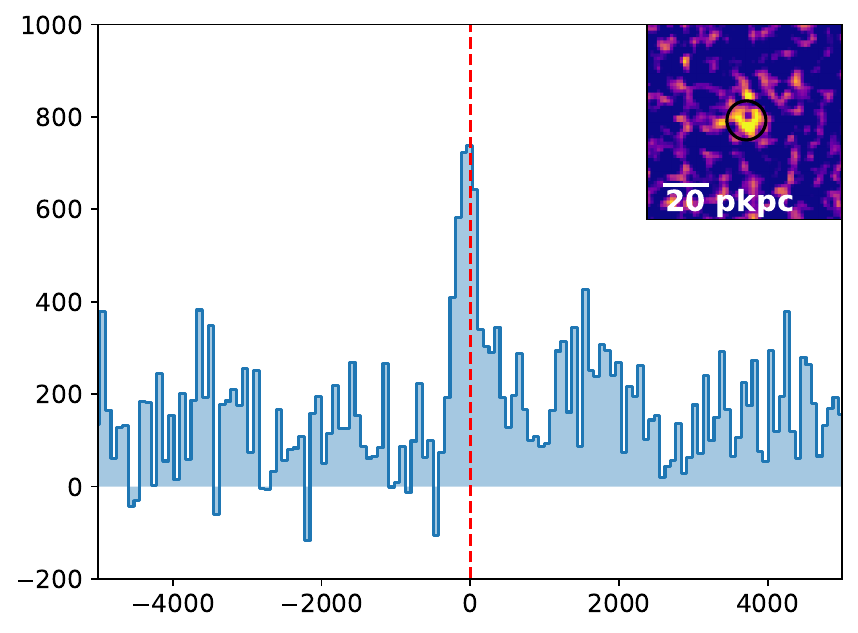}
    \includegraphics[width=.42\textwidth]{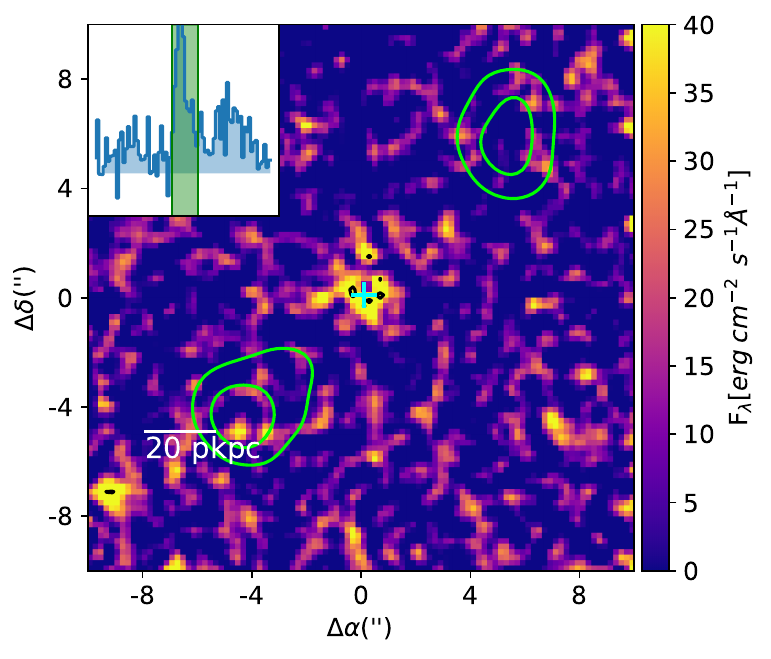}
    \includegraphics[width=.5\textwidth]{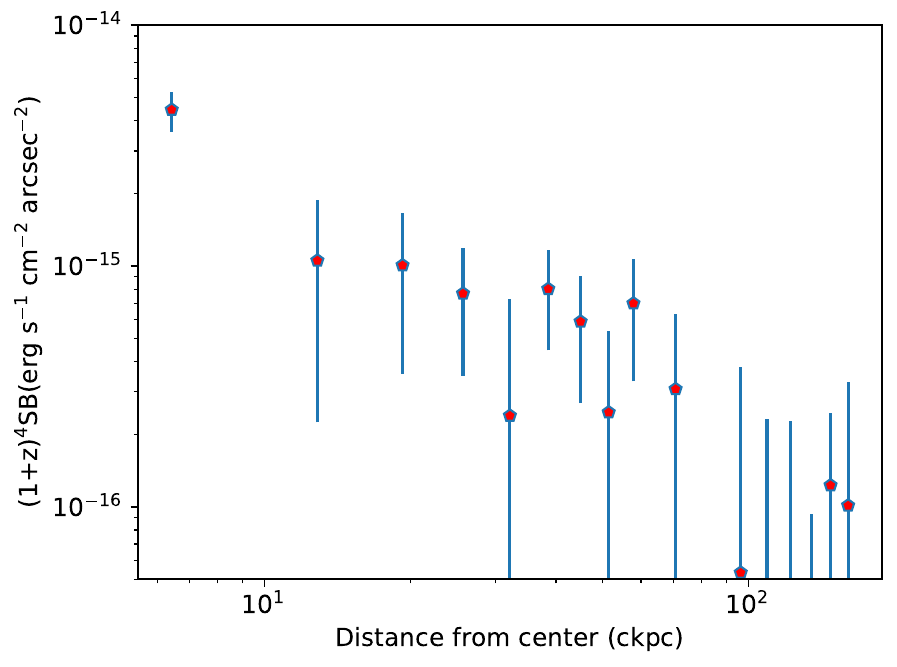}
    \caption{NVSS~J095751-213321. Top: nuclear spectra (left) and \lya\ narrowband image (right) extracted from the spectral region shown in the 10"$\times$10" inset with superposed radio contours obtained from the VLASS image. The black contours mark the 2$\sigma$ flux limits, and the blue cross is the point from which the brightness profile is extracted. Bottom: \lya\ brightness profile.}
    \label{fig:j095nuc}
\end{figure*}

\begin{figure*}
    \centering
    \includegraphics[width=\textwidth]{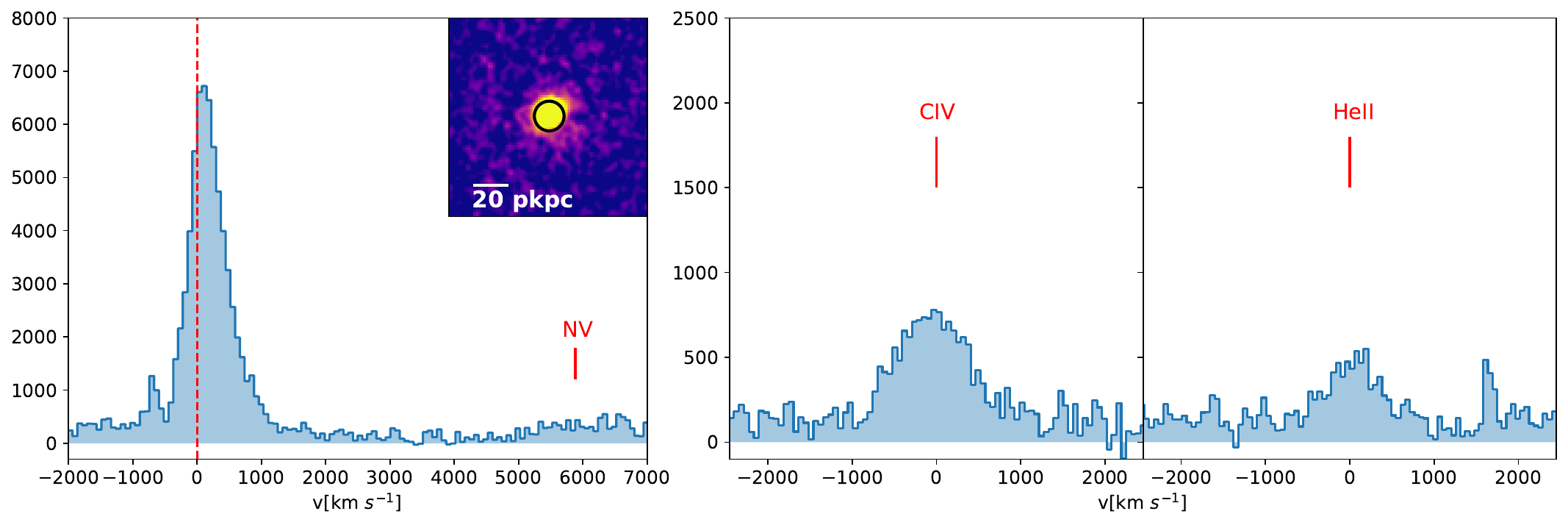}
     \includegraphics[width=0.5\textwidth]{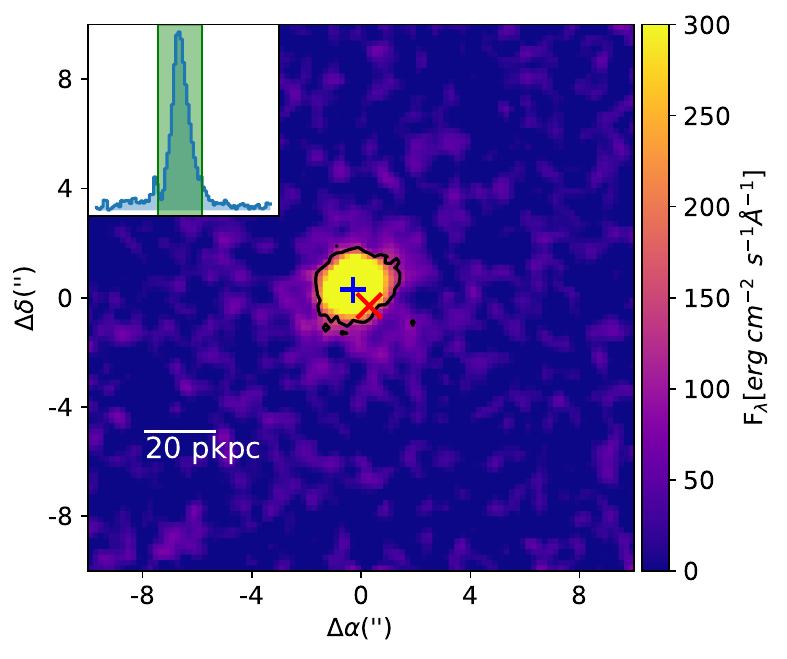}
\includegraphics[width=0.45\textwidth]{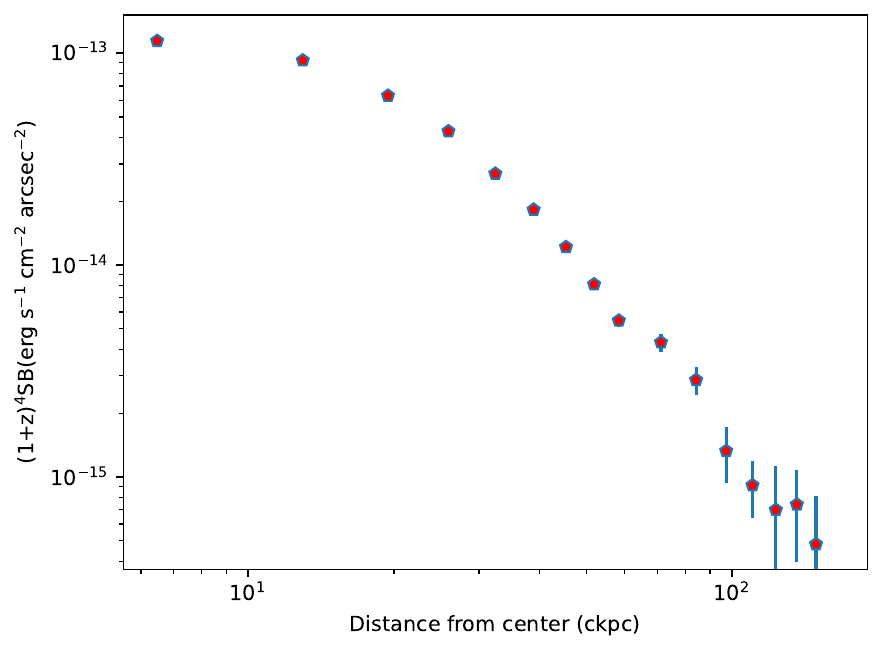}
\includegraphics[width=\textwidth]{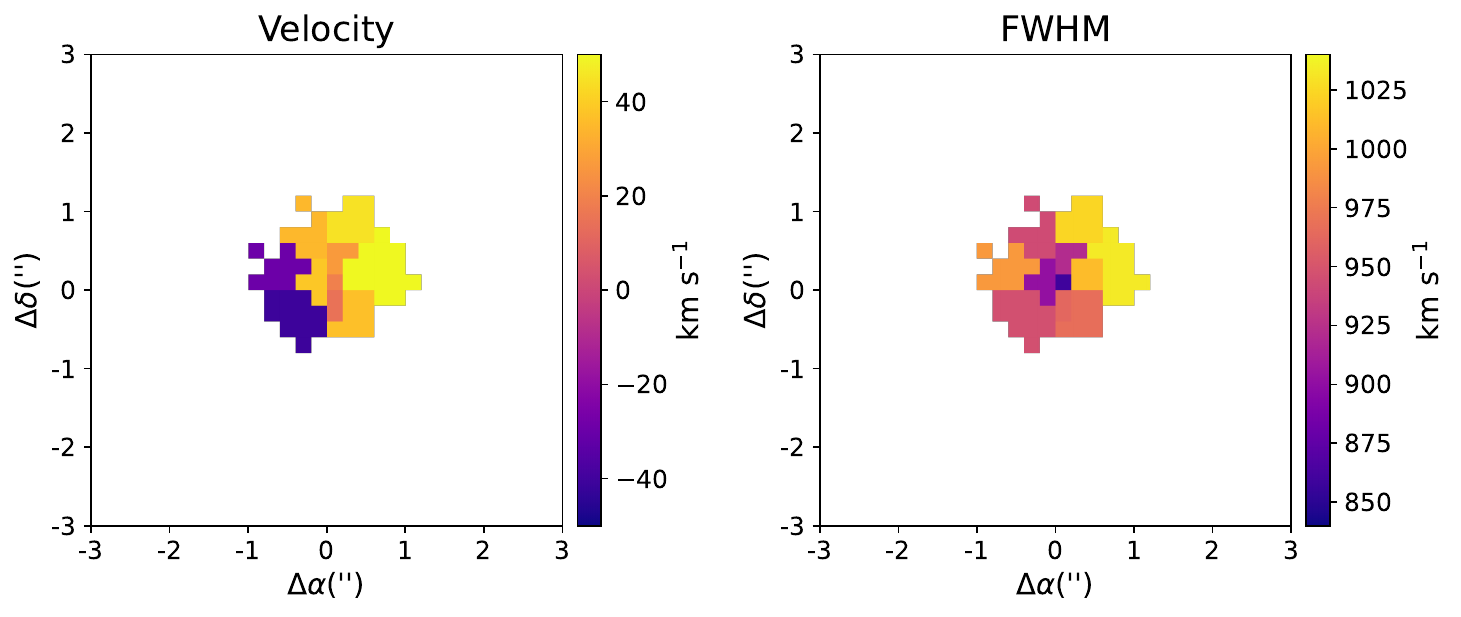}
   \caption{MRC~0251-273. Top: nuclear spectra centered on the \lya\ (left) and CIV + HeII lines (right) extracted from the region shown as black circle in the inset. Middle: left, \lya\ narrowband image extracted from the spectral region shown in the 10"$\times$10" inset; right: \lya\ brightness profile. The black contours mark the 2$\sigma$ flux limits, and the blue cross is the point from which the brightness profile is extracted. The location of the radio source, unresolved in the VLASS image, is marked with a red cross. Bottom: Voronoi-tessellated maps of velocity and FWHM of the \lya\ line.}
    \label{fig:ex2}
\end{figure*}

\begin{figure*}
    \centering
    \includegraphics[width=.52\textwidth]{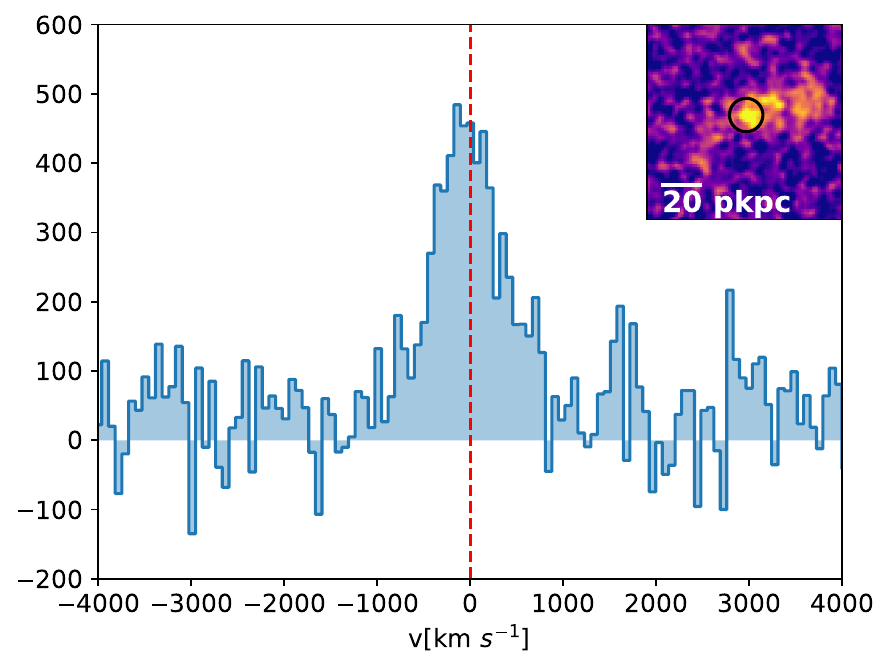}
    \includegraphics[width=.42\textwidth]{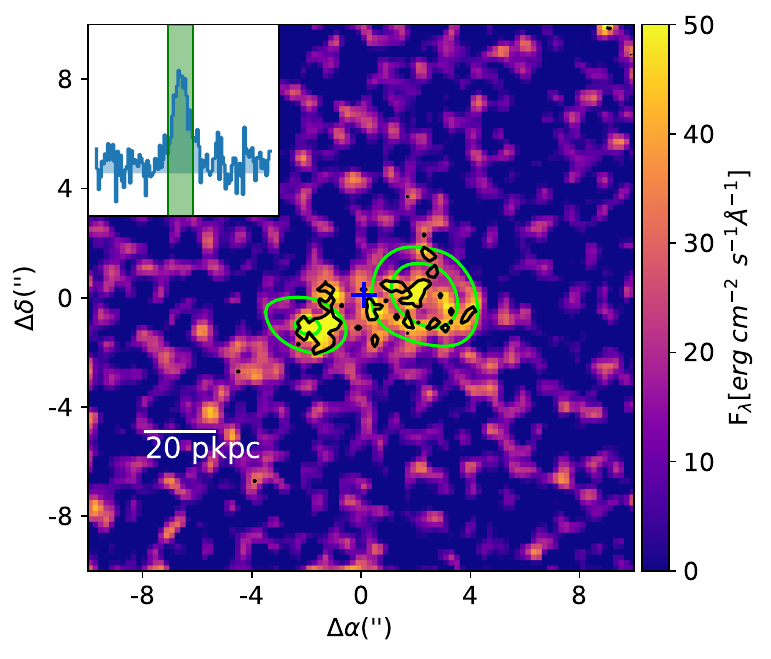}
    \includegraphics[width=.5\textwidth]{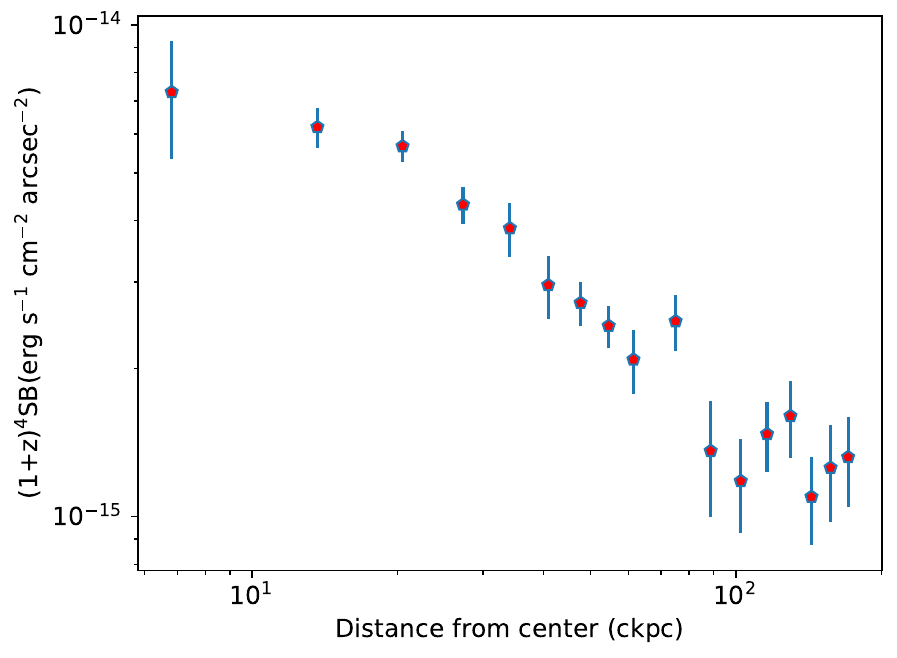} 
   \caption{NVSS~J094724-210505. Top: nuclear spectra (left) and \lya\ narrowband image (right) extracted from the spectral region shown in the 10"$\times$10" inset with superposed radio contours obtained from the VLASS image. The black contours mark the 2$\sigma$ flux limits, and the blue cross is the point from which the brightness profile is extracted. Bottom: \lya\ brightness profile.}
    \label{fig:E138nuc}
\end{figure*}

\begin{figure*}
    \centering
    \includegraphics[width=\textwidth]{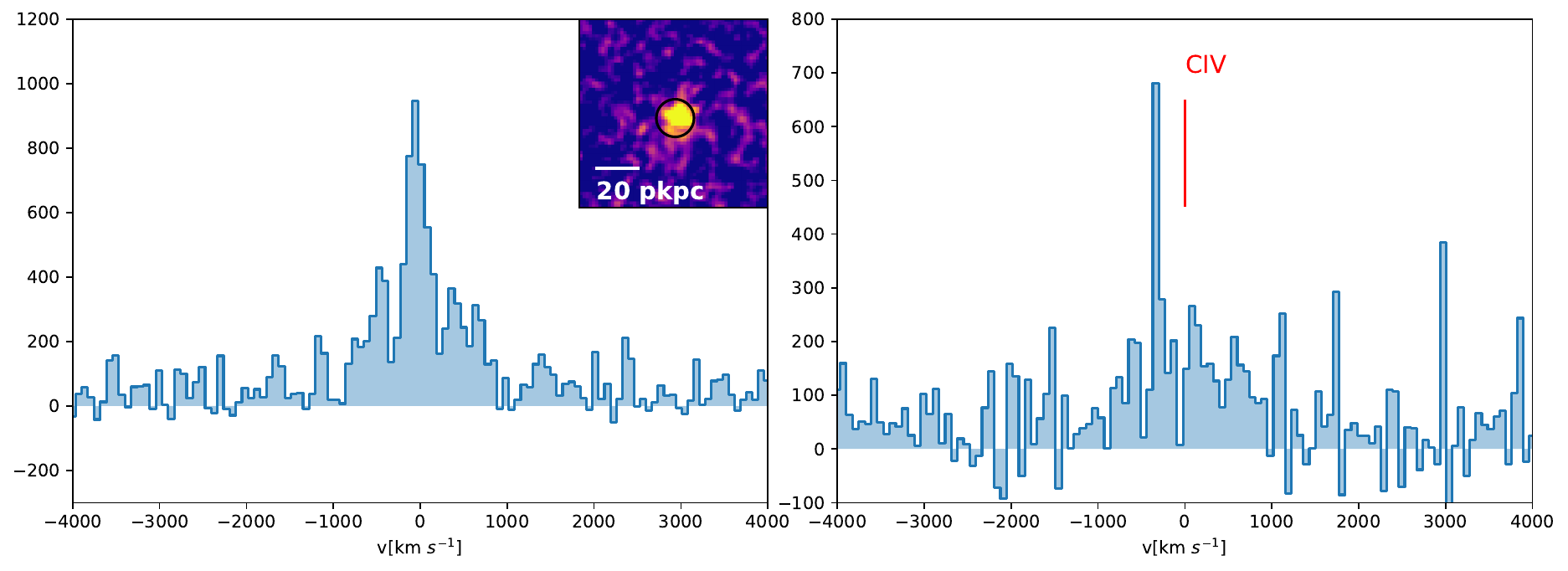}
    \includegraphics[width=.5\textwidth]{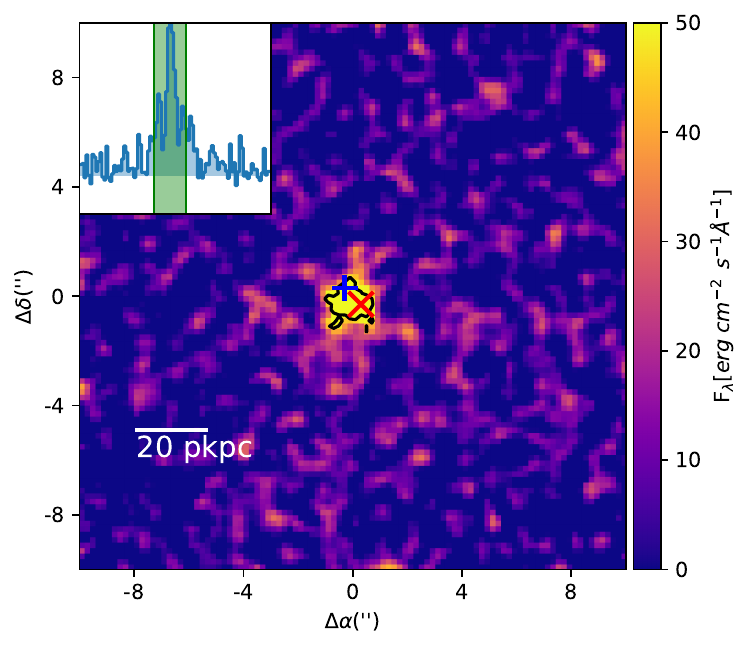}
    \includegraphics[width=.45\textwidth]{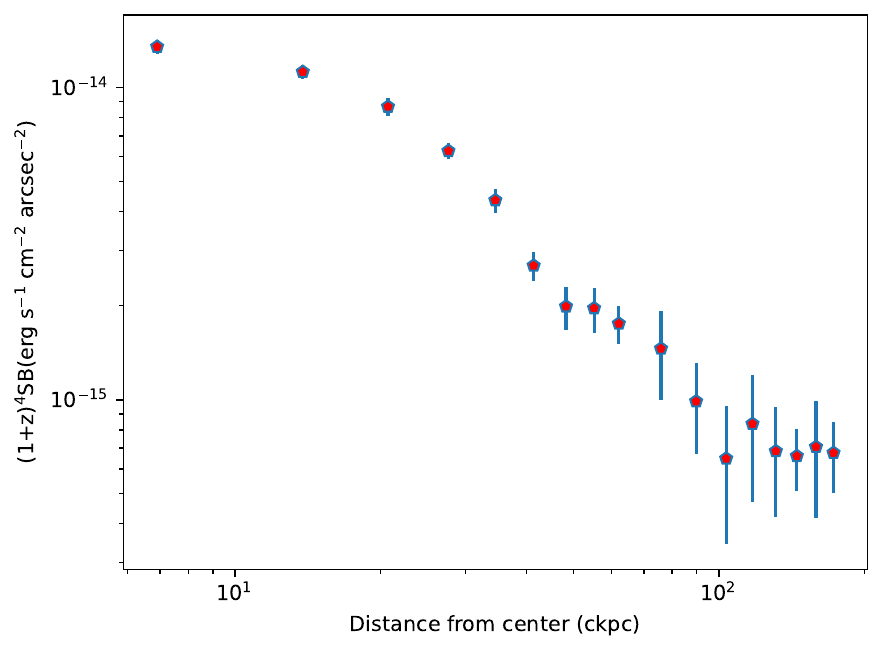}\\
    \includegraphics[width=\textwidth]{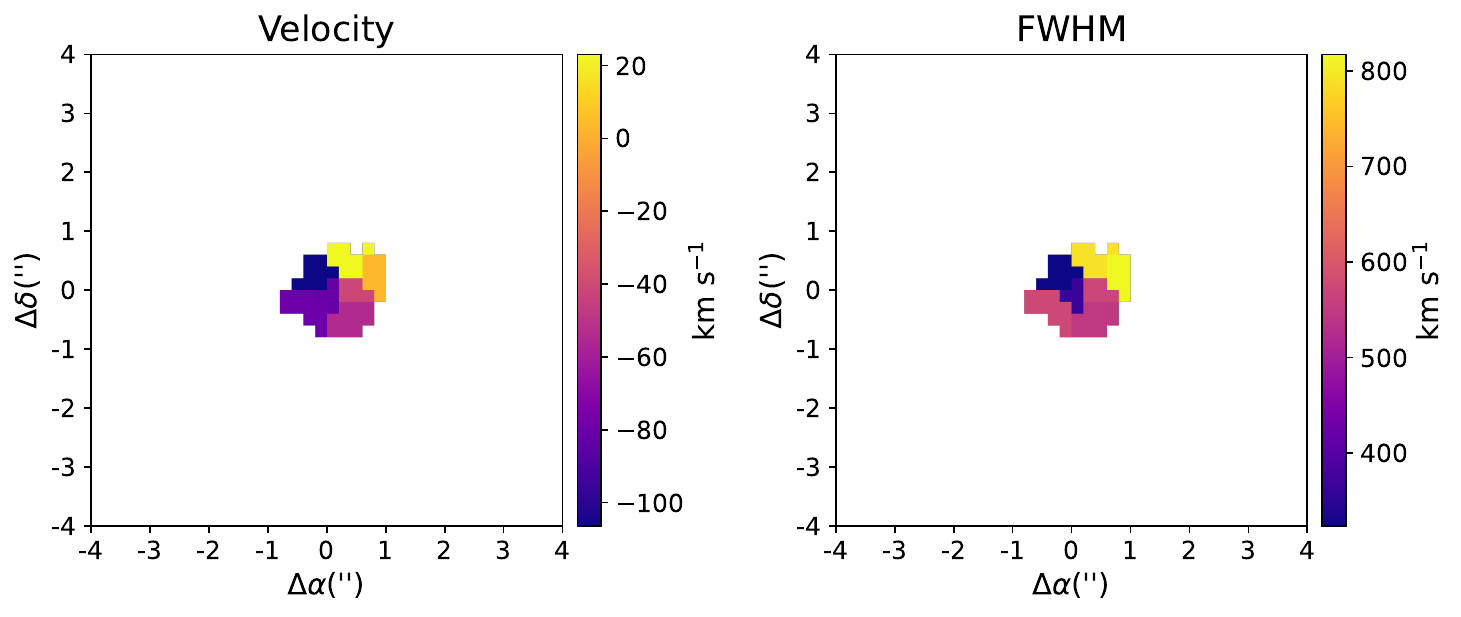}
    \caption{NVSS~J095438-210425. Top: nuclear spectra centered on the \lya\ (left) and CIV (right) extracted from the region shown as black circle in the inset. Middle: left, \lya\ narrowband image extracted from the spectral region shown in the 10"$\times$10" inset; right: \lya\ brightness profile. The black contours mark the 2$\sigma$ flux limits, and the blue cross is the point from which the brightness profile is extracted. The location of the radio source, unresolved in the VLASS image, is marked with a red cross. Bottom: Voronoi-tessellated maps of velocity and FWHM of the \lya\ line.}
    \label{fig:E32nuc}
\end{figure*}

\begin{figure*}
    \centering
    \includegraphics[width=.58\textwidth]{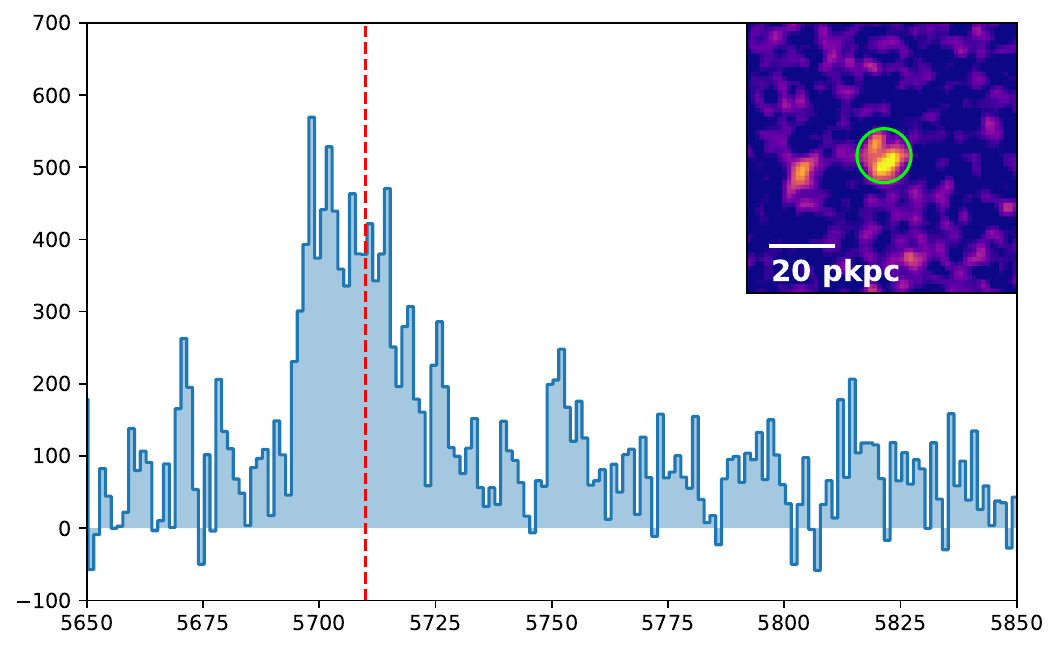}
    \includegraphics[width=.42\textwidth]{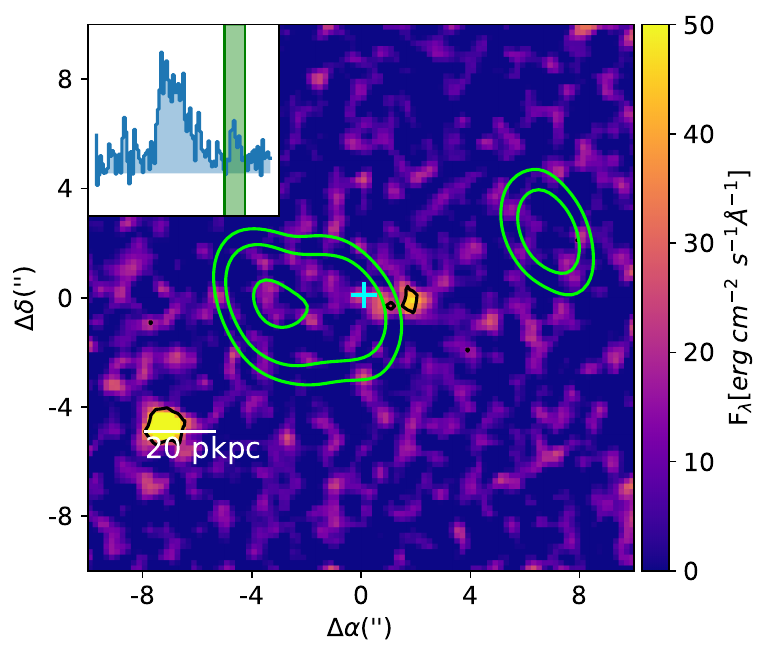}
    \includegraphics[width=.53\textwidth]{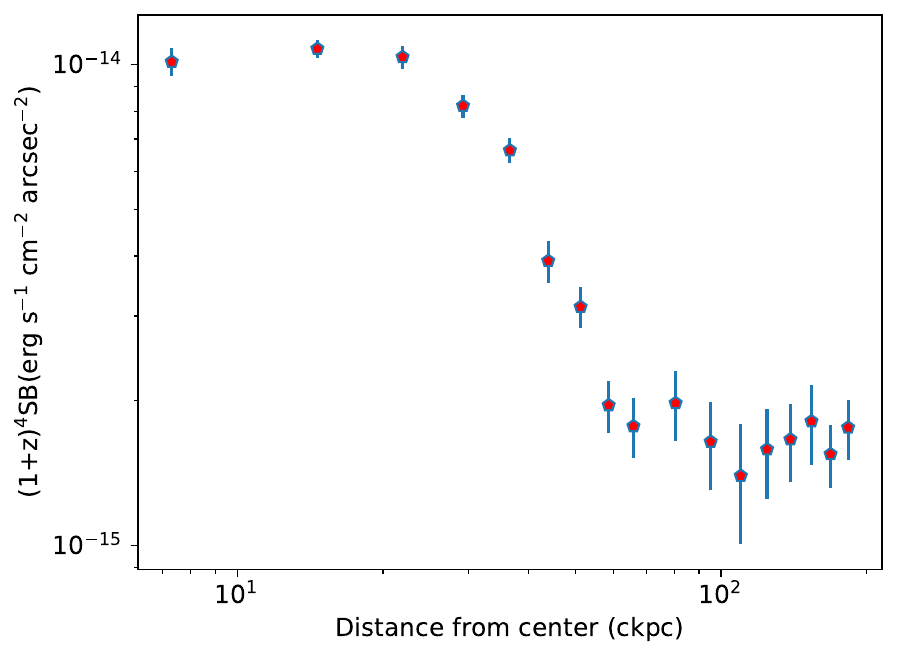}  
    \caption{TN~J1049-1258: Top: nuclear spectra centered on \lya\ (left) and CIV (right). Bottom left: narrow band image (left) with superposed radio contours obtained from the VLASS image. The black contours mark the 2$\sigma$ flux limits, and the blue cross is the point from which the brightness profile is extracted. Bottom right \lya\ brightness profile.}
    \label{fig:TNnuc}
\end{figure*}

\begin{figure*}
    \centering
    \includegraphics[width=.58\textwidth]{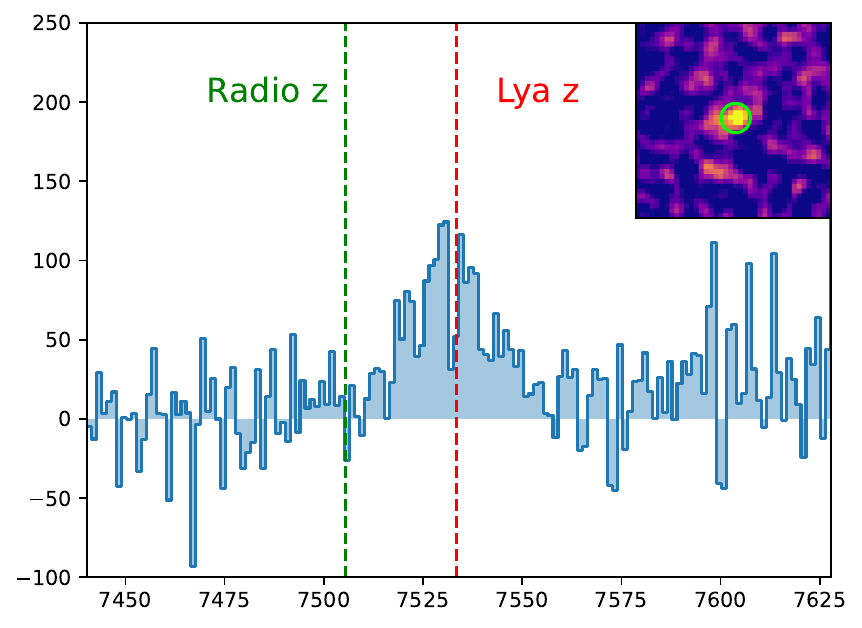}
    \includegraphics[width=.48\textwidth]{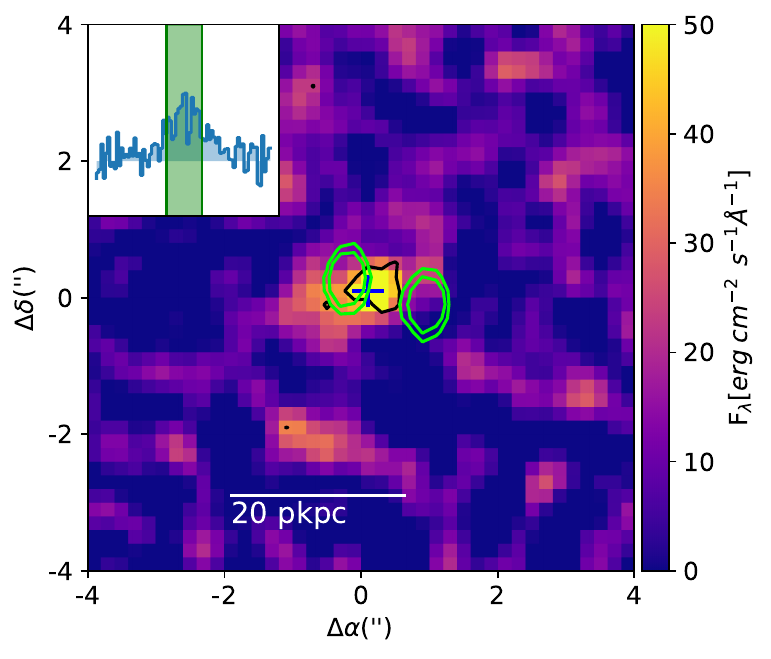}
    \includegraphics[width=.50\textwidth]{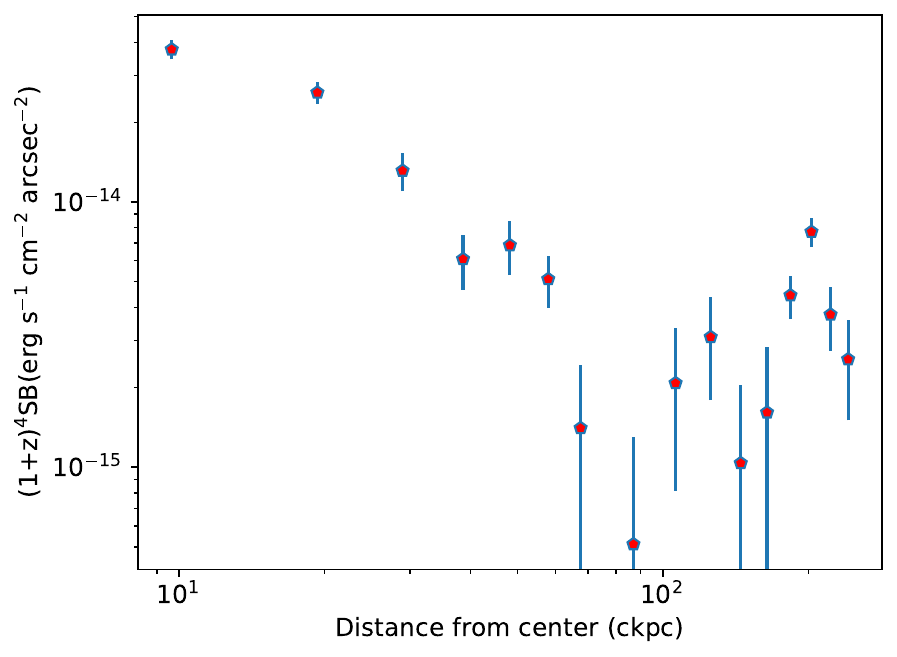}
    \caption{TN~J0924-2201: Top: nuclear spectra centered on \lya. The red dashed line corresponds to the redshift measured as the centre of the \lya\ emission, while the green line corresponds to the centroid measurement of the [CII]$\lambda$158$\mu$m molecular line \citep{lee2024ongoingfossillargescaleoutflows}. Bottom: narrow band image (left) and \lya\ brightness profile (right). The black contours mark the 2$\sigma$ flux limits, and the blue cross is the point from which the brightness profile is extracted. While the radio emission is unresolved in the VLASS image, a higher resolution image (FWHM = 0\farcs49) at 8.5 GHz shows two components separated by 1\farcs2, oriented at PA$\sim 70^\circ$.}
    \label{fig:j092nuc}
\end{figure*}

\clearpage
\newpage
\section*{Appendix B: Notes on the individual sources.}\label{appendixB}
\label{notes}

\textbf{- NVSS J151020-352803:} z=2.937. Radio morphology: asymmetric double lobe, extension of 8\farcs2\, with PA = -65$^\circ$. This source shows in its \lya\ emission a clear double lobed morphology, oriented in an east-west direction with a slight valley separating the two emission regions. The velocity difference between the \lya\ centroids is $\sim$120\kms. There are no additional visible lines on either region. The direction, size and position of the radio emission coincide with the observed \lya\ morphology, suggesting a relation between both emissions. There are various hypotheses which could explain the morphology of this source, for instance, a recent merger with another massive object which has strongly disturbed the galactic morphology. In this post-merger state, the relative velocities of both sources have come close to zero as they spiral towards one another. Another possibility would be the presence of a very dense region of neutral HI gas along the line of sight, located in the center of the source, absorbing \lya\ photons and creating the illusion of a double lobe morphology, or possibly a dust lane. Given the shape of the emission lines, there are clear signatures of absorption in the spectra of both lobes, so this is a strong possibility as well. However, the spatial correlation between radio and ionized emission suggests a nuclear origin for the large scale morphology of the \lya\ emission. This final hypothesis could very well be connected to a prior merger, as it is a well known AGN triggering mechanism. It remains unclear which of these hypotheses best explains our observations. 

\textbf{- TXS 0952-217:} z=2.942$\pm$0.001 from HeII$\lambda$1640. Radio morphology: point source, upper limit to its deconvolved size is 1.2". The second-brightest source in our sample, TXS 0952-217 presents a highly absorbed \lya\ profile that changes along the spatial extension of the nebula. This source was the object of study in \citet{puga202420}, where an extended outflow of ionized gas was found via modeling of the observed emission profile. CIV, HeII, and CIII] are also observed, obtaining a robust redshift measurement as well as several line ratios, which were consistent with an AGN origin, as no clear signs of enhanced star formation were observed. While both the galaxy and outflow are extended ($\sim$25 kpc for the galaxy and $\sim$20 kpc for the outflow), the radio source is point-like. The kinematics of the host do not suggest orderly motion, but this might be the result of a very limited S/N.

\textbf{- TN J1112-2948:} z=3.089$\pm$0.001 from HeII$\lambda$1640. Radio morphology; double lobe, extension of 10\arcsec, PA = -45$^\circ$. This object presents a heavily absorbed \lya\ profile, as is made apparent by the various peaks and valleys on the red side of the line, and the extremely narrow emission peak given how wide the total line profile is. CIV and HeII are also detected. The \lya\ morphology is quite complex in this object. It is the only object beside TN 1049-1258 for which we confirmed the presence of a close companion galaxy, $\sim$30 kpc southeast of the host. In the companion we detect the same ionized lines as in the host, although the detection of CIV is borderline. Both the narrow width of the lines and the line ratio between \lya\ and HeII indicate this companion is a quiescent galaxy displaying no signs of nuclear activity. In addition, there is a filament of ionized gas north of the host galaxy, which is only visible in \lya\ emission. The very elongated morphology of this emission suggests it may be part of the extended nebula instead of an additional companion, as well as the fact that no helium or carbon lines are detected. There is no detectable continuum emission on any of the sources, likely caused in part by contamination due to the presence of a magnitude 7 star, HD 97412, slightly outside the northwest corner of the image. 

\textbf{- NVSS J095751-213321:} z=3.126. Radio morphology: double lobe, extension of 16\arcsec, PA = -45$^\circ$. Only a small region of \lya\ emission is detected in this source. The morphology of the \lya\ emission appears clumpy and very faint, almost certainly the result of resonant scattering. The emission is extremely narrow (FWHM=325$\pm$43\kms) compared to the typical widths for RGs at this redshift, and by far the narrowest in our sample or W23+, which confirms the presence of intervening gas along our line of sight. No other lines are observed and no companions are found in the MUSE field. The radio emission, which displays a very large double lobe in the SE-NW direction, appears to be entirely disconnected from the ionized gas, although deeper observations will be necessary in order to ascertain the true extent of the nebula. 

\textbf{- MRC 0251-273:} z=3.163$\pm$0.001 from the He line. Radio morphology: point source. The brightest object in our sample, it displays relatively compact and symmetric \lya emission, and very compact NV, CIV and HeII emission. This object displays a high \lya/HeII flux ratio of $\sim$16; values this high have been previously associated to enhanced star formation \citep{Villar_Mart_n_2007_a}. The kinematic analysis of the \lya\ emission yields widths in the order of $10^{3}$\kms, and its velocity field is suggestive of a rotating disk.

\textbf{- NVSS J094724-210505:} z=3.374$\pm$0.001 from \lya. Radio morphology: asymmetric double lobe, extension of 4.1", PA = -76°. From its \lya\ emission, this object appears to be the largest among those observed in this sample, with a projected extension of about 70 kpc in the E-W direction. Similarly to NVSS J151020-352803, the radio and \lya\ emission appear to be cospatial, although the true extent of the ionized nebula is difficult to parse given how faint it appears. No other lines are observed in its spectrum, and the width of the \lya\ profile is within normal bounds for HzRGs (FWHM=870$\pm$90 \kms).

\textbf{- NVSS J095438-210425:} z=3.431. Radio morphology: point source. Relatively compact and faint \lya emission. Out of the other ionized lines, only CIV is visible; as it is also a resonant line as well as a doublet, it does not necessarily yield a more robust systemic redshift measurement. The rotation of the ionized gas is clear in spite of its intrinsic faintness, suggesting that the observed emission corresponds to a galactic disk. The line emission appears to be quite narrow when compared to the rest of the sample, likely due to intervening gas, which is visible in the many absorption troughs present in the nuclear spectrum. Given the upper bound of HeII flux, which appears to be right under the detection threshold, CIV seems to be at minimum 4 times brighter. This is an outlier in our sample, as we generally observe the CIV/HeII to hover close to 1.

\textbf{- TN J1049-1258:} z=3.697. Radio morphology: asymmetric double lobe, extension of 10", PA is -80°. This source consists of what appear to be two distinct patches of \lya\ emission. The RG itself, identified by its position relative to the radio emission as well as by virtue of being the brightest emission line source, only presents \lya\ emission. Slightly displaced both in the spatial and frequency domains, there is an extended patch of emission west of the RG and aligned with the radio axis, for which \lya\ and HeII are visible. There are various possibilities for the origin of this emission, including a companion galaxy and an outflow of ionized gas. The source is also embedded within what appears to be a protocluster, with two other separate \lya\ emission galaxies situated at 60 pkpc and 25 pkpc. A deeper study of both the nature of the extended emission as well as the environment of this source was performed in \citet{puga2023extended}.

\textbf{-TN J0924-2201:} z=5.197. Radio morphology: double source, $\arcsec 1.4$ in size at PA 75. At a redshift of $\sim$5.2 it is the farthest source in our sample, and one of the highest redshift HzRGs found to date. \lya\ is barely detectable by integrating the whole emission of the galaxy. It appears very compact, as it is the only nebula which cannot be spatially resolved due to being smaller or around the size of the seeing. The line profile is in agreement with what is generally observed in HzRGS (FWHM=880$\pm$170 \kms). \citet{lee2024ongoingfossillargescaleoutflows} detected high S/N [CII]$\lambda$158$\mu$m emission which yielded a redshift of z=5.1736$\pm$0.0002, corresponding to an offset of $\sim$10$^{3}$\kms\ between \lya\ and the systemic redshift, the largest observed to date. This suggests that the \lya\ emission from this source is extremely absorbed given that, both in our observations with MUSE and those performed by \citet{Matsuoka_2011} with FOCAS at Subaru \citep{2002kashikawa}, no flux is detectable at the theoretical line center of \lya. The actual ionized emission from this galaxy is likely much higher, but the low S/N of the ionized line observations, along with the implicit degeneracies between HI column density and intrinsic line flux, make it impossible to ascertain the extent of the effects of resonant scattering given the currently available data. Companion CO(1–0) has been detected from galaxies close to this source \citep{Lee_2023}, but none of these emitters are visible in the MUSE image, neither in continuum nor \lya. CIV was also detected by \citealt{Matsuoka_2011}, but, along with HeII and CIII], falls outside of our spectral coverage with MUSE given the object's very high systemic redshift.

\end{appendix}

\section*{Appendix C: Undetected sources}\label{appendixC}

In two of 11 sources observed, \lya\ emission was not detected anywhere within the MUSE data. They are NVSS J021308-322338 (z=3.976$\pm$0.001,  identified by \citealt{de_Breuck_2004}) and TN J1123-2154 (z=4.109$\pm$0.004, identified by \citealt{De_Breuck_2001}). In order to obtain an upper bound to their fluxes, we placed two synthetic apertures with a radius of 1\farcs2\ at the nominal centers of the radio structure. The spectra extracted at these locations are shown in Fig. \ref{fig:nondetec}. 

For NVSS J021308-322338, the flux reported by \citealt{de_Breuck_2004} is (0.16$\pm$0.02)$\times$10$^{-16}$ erg s$^{-1}$ cm$^{-2}$. The 3$\sigma$ upper bound to the \lya\ flux (centered at the wavelength of the reported detection) we obtain for this source based on the MUSE data is $F_{\rm Ly \alpha}$ < 0.08$\times$10$^{-16}$ erg s$^{-1}$ cm$^{-2}$. This measurement has been obtained adopting a conservative value for the integration range in wavelength corresponding to 1,000 \kms. For TN J1123-2154, the 3$\sigma$ upper bound is 0.07$\times$10$^{-16}$ erg s$^{-1}$ cm$^{-2}$, whilst the flux observed by \citealt{de_Breuck_2004} is (0.18$\pm$0.05)$\times$10$^{-16}$ erg s$^{-1}$ cm$^{-2}$. In conclusion, we can not confirm the identification of these two sources as HzRGs. 

\renewcommand{\thefigure}{C.\arabic{figure}}
\setcounter{figure}{0}

 \begin{figure}[ht!]
   \centering
    \includegraphics[width=.50\textwidth]{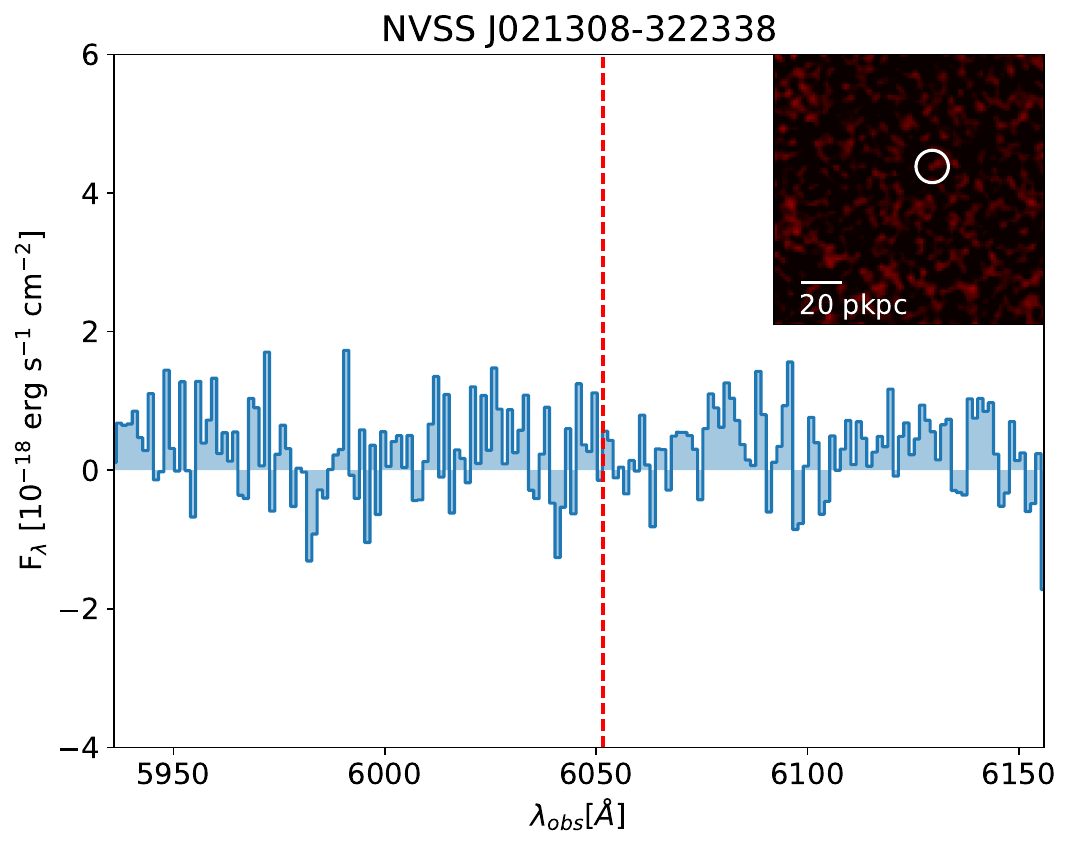}\\
    \includegraphics[width=.50\textwidth]{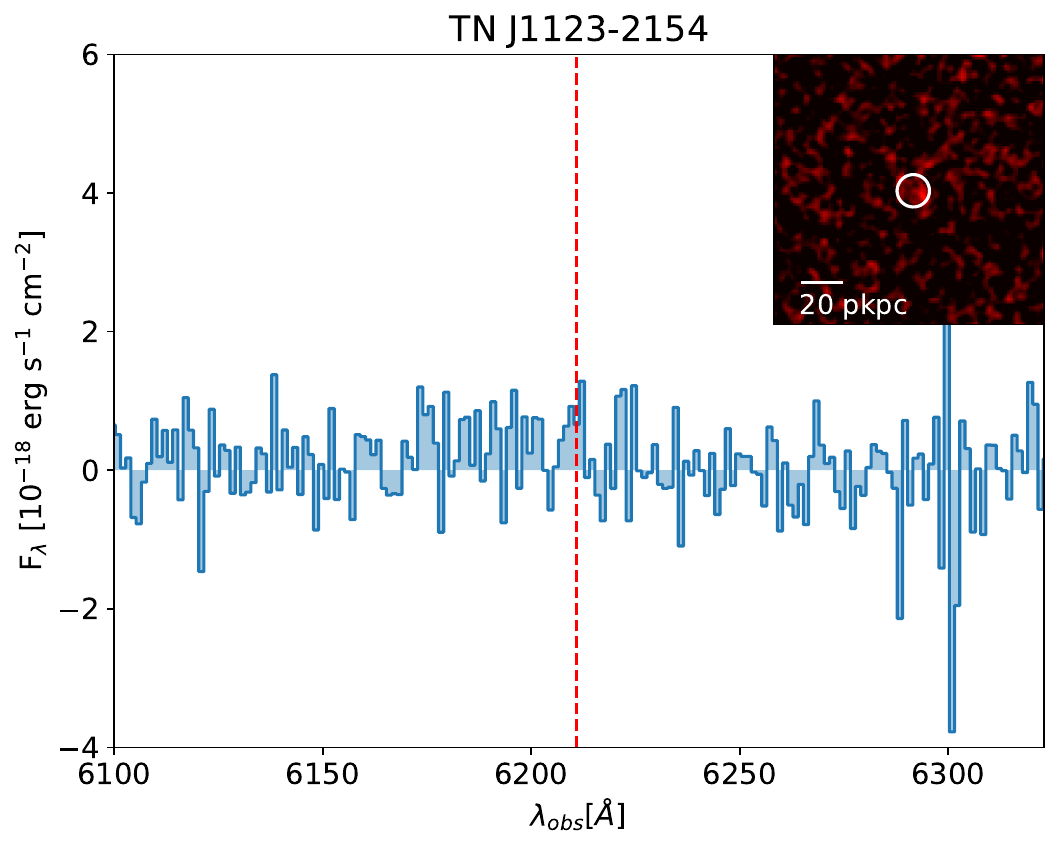}
    \caption{Nuclear spectra of NVSS J021308-322338 (left) and TN J1123-2154 (right) extracted using a circular synthetic aperture with a radius of 1\farcs2\ located at the center of their radio emission. The vertical dashed red lines mark the wavelength of the reported \lya\ detection. No line emission is visible in these spectra. The inset on the top right corresponds to a narrowband image extracted with a width corresponding to 1,000 \kms.}
     \label{fig:nondetec}
\end{figure}

\end{document}